\documentclass[aps,prb,twocolumn,showpacs,amsmath,amssymb,superscriptaddress,floatfix]{revtex4-1}
\usepackage{graphicx}
\usepackage{dcolumn}
\usepackage{bm}
\usepackage{amsfonts}
\usepackage{amsmath}
\usepackage{ulem}
\usepackage{color}
\usepackage{hyperref}

\begin{document}

\title{Interplay between Rashba interaction and electromagnetic field \\in the edge states of a 2D~topological~insulator}

\author{Fabrizio Dolcini}
\email{fabrizio.dolcini@polito.it}
\affiliation{Dipartimento di Scienza Applicata e Tecnologia del Politecnico di Torino, I-10129 Torino, Italy}
\affiliation{CNR-SPIN, Monte S.Angelo - via Cinthia, I-80126 Napoli, Italy}

\begin{abstract}
The effects of Rashba interaction and electromagnetic field on the edge states of a two-dimensional topological insulator are investigated in a non-perturbative way. We show that the electron dynamics is equivalent to a problem of massless Dirac fermions propagating with an inhomogeneous velocity, enhanced by the Rashba profile with respect to the bare Fermi value $v_F$. Despite the inelastic and time-reversal breaking processes induced by the electromagnetic field, no backscattering occurs without interaction. The photoexcited electron densities are  explicitly obtained in terms of the electric field and the Rashba interaction, and are shown to fulfil  generalised chiral anomaly equations. The case of a Gaussian electromagnetic pulse is analysed in detail. When the photoexcitation occurs far from the Rashba region, the latter effectively acts as a ``superluminal gate''   boosting the photoexcited wavepacket outside the light-cone determined by $v_F$. In contrast, for an electric pulse overlapping the Rashba region the emerging wavepackets are squeezed in a manner that depends on the overlap area. The electron-electron interaction effects are also discussed, for both intra-spin and inter-spin density-density coupling. The results suggest that Rashba interaction, often considered as an unwanted disorder effect, may be exploited to tailor the shape and the propagation time of photoexcited spin-polarised wave packets. 

\end{abstract}


\maketitle
\section{Introduction}
One of the most fascinating properties of Topological Insulators (TI) is the helical nature of their edge states, namely the locking of the propagation direction to the spin orientation~[\onlinecite{qi-zhang_2011,hasan_2010,ando_2013}], which makes them a promising platform for spintronics applications~[\onlinecite{moore_2010,qi-zhang_2010,macdonald_2012,murakami_2014}]. In this respect, an exciting perspective is the possibility to photoexcite spin-polarised electron wave packets by coupling the edge states to an external electromagnetic radiation. Notably, the very helicity of the edge states, which is closely related to their  massless Dirac fermions behavior, prevents  electron photoexcitation  via the customary vertical electric dipole transitions that occur in optoelectronics of conventional semiconductors. 
To overcome this problem, strategies have been proposed,  particularly for two-dimensional time-reversal symmetric  TIs, i.e. quantum spin Hall (QSH) systems~[\onlinecite{kane-mele2005a,kane-mele2005b,bernevig_science_2006,konig_2006,molenkamp-zhang_jpsj,roth_2009,brune_2012,liu-zhang_2008,knez_2007,knez_2014,spanton_2014}].
A circularly polarised radiation, for instance, can induce magnetic dipole transitions via Zeeman coupling to the edge states~[\onlinecite{kindermann2009,cayssol2012,artemenko2013,dolcetto-sassetti2014}]. Its efficiency, however, is limited by the  $g$-factor.  
Alternatively, exploiting frequencies that exceed the bulk gap, optical  transitions can be induced from the edge states to the bulk states~[\onlinecite{artemenko2013},\onlinecite{artemenko2015}], which, however, are not topologically protected.  

Quite recently it has been shown that, when an electromagnetic field is applied on a spatially localised region and for a finite time, electron photoexcitation consisting of purely intra-branch transitions can be induced on the edge states without invoking the bulk states or the Zeeman coupling~[\onlinecite{dolcini_2016}]. 
Localised electromagnetic pulses can be generated either by near field spectroscopy~[\onlinecite{novotny_review,koch1997,novotny2003,nomura2011,nomura2015}], or by  thin metallic electrodes coupled to a  QSH edge and biased by a time-dependent voltage~$V(t)$~[\onlinecite{levitov_2006,glattli_PRB_2013,glattli2013,bocquillon2014}].  As a result of the localised photoexcitation process, spin-polarised electron wave packets are thus predicted to propagate along the edge, maintaining their shape without dispersion~[\onlinecite{dolcini_2016}]. 

However, an important aspect of the wave packet photoexcitation  and propagation process has not been discussed so far, namely the interplay between the electromagnetic radiation and the Rashba interaction along the edge. The Rashba interaction, which couples electrons of opposite spins and momenta, is due to a local inversion asymmetry originating by mainly two mechanisms. On the one side it may be the result of accidental disorder, caused for instance by the random ion distribution in the heterostructure doping layers or by the presence of random bonds at the  quantum well interfaces, e.g. due to the etching process~[\onlinecite{sherman_PRB_2003},\onlinecite{sherman_APL_2003}]. On the other hand, local strain and Rashba interaction can be generated by deforming the curvature of the geometrical boundaries~[\onlinecite{entin_2001,ojanen_2011,ortix_2015,gentile_2015,cuoco-2016}], or by  applying an electric field to  metallic gate electrodes~[\onlinecite{molenkamp_2006,niu_2013,park_2013,wolosyn_2014}]. In these cases, it may   be significantly strong and cannot be treated as a weak perturbation. It is worth stressing that the Rashba interaction --no matter how strong-- cannot lead, alone,  to any electron backscattering. This is because it is an elastic process that preserves time-reversal symmetry. Such  topological protection argument, however, cannot be invoked in the presence of inelastic and/or time-reversal breaking processes. When  Rashba interaction interplays with electron-phonon~[\onlinecite{budich_2012}] or with electron-electron interaction~[\onlinecite{johannesson_2010,
crepin_2012,schmidt_2012,geissler_2014,mirlin_2014,geissler_2015}], for instance, electron backscattering can arise in the edge states.
 
In this paper we focus on the helical edge states of a two-dimensional TI, and investigate non-perturbatively the interplay between Rashba interaction and the electromagnetic field, which induces inelastic and time-reversal breaking processes. We show that, via an appropriate rotation of the electron field operator into a local chiral basis, the problem can be recast into a massless Dirac fermion dynamics, characterised by an inhomogeneous velocity. This mapping enables us to obtain exact expressions for the dynamical evolution of photoexcited electron wave packets, unveiling two important features. In the first instance, no backscattering arises as long as electron-electron interaction can be neglected,   despite the inelastic scattering and the break up of time reversal symmetry due to the electromagnetic field.
Secondly, the velocity renormalised  by   the Rashba interaction  turns out to be always increased with respect to the bare Fermi velocity $v_F$, regardless of the sign and magnitude of the Rashba interaction. This entails interesting consequences for the electron photoexcitation problem. In particular, when the electromagnetic field is localised away from the Rashba coupling region, the Rashba region effectively acts a ``superluminal gate'': the photoexcited wave packet reaching the Rashba region is boosted to space-time regions lying outside the light-cone determined by $v_F$. In contrast, when the electromagnetic field overlaps the Rashba region, the photoexcited wave packets  emerging from it are squeezed and their shape depends on the overlap with the applied electromagnetic pulse. These results suggest that the Rashba interaction, often regarded to as an unwanted disorder effect, can be exploited also as a useful tool to tailor the shape and the propagation timescale of photoexcited wave packets.

The article is organised as follows. In Sec.~II  we present the model and its mapping into a model of Dirac electrons with an inhomogeneous velocity. In Sec.~III we discuss the effects of the Rashba interaction, showing  the behavior of the electron wave function impinging on a Rashba region.  In Sec.~IV  we include the electromagnetic field and explicitly provide the solution of the full problem, with a focus on the photoexcited electron densities and the chiral anomaly due to  Rashba interaction. In Sec.~\ref{sec-5} we apply the general results previously obtained to the specific example of a Gaussian electric pulse applied in the presence of a Rashba barrier, analyzing in detail various situations, from the case where the photoexcitation occurs away from the Rashba interaction to the case where an overlap  exists. Finally, in Sec.~\ref{sec-6} we discuss the effects of Zeeman coupling and electron-electron interaction, considering both intra-spin and inter-spin density-density coupling  and specifying the conditions under which the topological protection is robust. After outlining some possible experimental realisation setups, we draw our conclusions in Sec.\ref{sec-7}.
 
\section{Model}
\label{sec-2} 
\subsection{Hamiltonian and equation of motion}
A Kramers' pair of one-dimensional helical states counterflowing along an edge~$x$ of a two-dimensional TI are denoted by $\uparrow$ and $\downarrow$, and described by a spinor field operator   
\begin{equation}\label{Psi-def}
\Psi(x)=\left(\begin{array}{l}\psi^{}_{\uparrow}(x) \\ \\ \psi^{}_{\downarrow}(x) \end{array} \right)\quad.
\end{equation}
The electronic system, exposed to both a Rashba interaction and to an electromagnetic field, is modelled by the Hamiltonian $\hat{\mathcal{H}}  = \hat{\mathcal{H}}_\circ\,\,+\, \hat{\mathcal{H}}_{em}$, where $\hat{\mathcal{H}}_\circ=\hat{\mathcal{H}}_{kin}+\hat{\mathcal{H}}_R$ contains the linear dispersion term,
\begin{equation}\label{Hkin}
\hat{\mathcal{H}}_{kin} = \int   dx \, \Psi^\dagger(x) \,  v_F \sigma_3  {p}_x \, \Psi(x)  
\end{equation}
with ${p}_x=-i\hbar \partial_x$, and the Rashba interaction 
\begin{eqnarray}
\label{HR}
\hat{\mathcal{H}}_R = \frac{1}{\hbar} \int   dx \, \Psi^\dagger(x) \,   \frac{1}{2} \left\{ \alpha_R(x) \, , {p}_x   \right\}  \sigma_2\, \Psi(x) 
\end{eqnarray}
characterised by a profile $\alpha_R(x)$. Here $\sigma_{1,2,3}$  are  Pauli matrices.
Furthermore,  the term
\begin{eqnarray}
\hat{\mathcal{H}}_{em} &= &{\rm e} \int   dx V(x,t)  \, \hat{n}\, -\frac{{\rm e}}{c} \int   dx \, A(x,t) \,\hat{J}\label{Hem} 
\end{eqnarray}
describes the electromagnetic coupling, where
\begin{equation}\label{n-def}
\hat{n}=\Psi^\dagger(x) \Psi(x) = \hat{n}_\uparrow+\hat{n}_\downarrow
\end{equation}
denotes the electron density coupled to the scalar potential $V(x,t)$, 
\begin{eqnarray}
\hat{J}&=&v_F \Psi^\dagger(x) \,   \left(\sigma_3+\frac{\alpha_R(x)}{\hbar v_F }   \sigma_2\, \right)\Psi(x)= \nonumber\\
&=& v_F\left[ \hat{n}_\uparrow-\hat{n}_\downarrow +\frac{i  \alpha_R(x)}{\hbar v_F}(\psi^\dagger_\downarrow\psi^{}_\uparrow-\psi^{\dagger}_\uparrow\psi^{}_\downarrow) \right] \label{J-def}
\end{eqnarray}
is the current density operator coupled to the vector potential $A(x,t)$, and $\hat{n}_{\uparrow(\downarrow)}\doteq \psi^\dagger_{\uparrow(\downarrow)}\psi^{}_{\uparrow(\downarrow)}$.  Notably, the expression (\ref{J-def}) of the current operator is affected by the Rashba profile $\alpha_R(x)$ in order to ensure charge conservation, as can straightforwardly be seen by applying the minimal coupling $\hat{p}_x\rightarrow \hat{p}_x-{\rm e}\, A/c$ to both terms of~$\hat{\mathcal{H}}_\circ$. The applied electric field is $E(x,t)=-\partial_x V- \partial_t A/c$, while the Zeeman coupling ascribed to the magnetic field arising from time-dependence of $E(x,t)$ is neglected.  
 The Hamiltonian $\hat{\mathcal{H}}$ can be compactly rewritten as
$
\hat{\mathcal{H}}  =  \int dx \, \Psi^\dagger(x) \, H(x) \, \Psi(x) \label{H}
$, 
where the  first-quantized Hamiltonian density is
\begin{eqnarray}
\lefteqn{H(x)  = } \label{Hmat} \\
&=&  \frac{v_F}{2} \left\{ \sigma_3 +\tan\theta_R(x) \sigma_2 \, ,  {p}_x - \frac{{\rm e}}{c} A(x,t) \right\} \, \, \, +{\rm e} V(x,t) \sigma_0\,\,, \nonumber
\end{eqnarray}
with $\sigma_0$ denoting the $2 \times 2$ identity matrix and $\theta_R \in [-\pi/2; + \pi/2]$  the Rashba angle, defined as
\begin{equation} \label{thetaR-def}
\theta_R(x) \doteq \arctan \frac{\alpha_R(x)}{\hbar v_F}  \quad.
\end{equation}
Equation (\ref{Hmat}) dictates the dynamical evolution $i \hbar\partial_t \Psi=  H\, \Psi$ of the electron field spinor. Without Rashba interaction ($\theta_R = 0$) and electromagnetic coupling ($A=V=0$) the equation of motion reduces to the customary helical form $\partial_t \Psi=-v_F \sigma_3 \partial_x \Psi$, and the two spin components $\psi_\uparrow$ and $\psi_\downarrow$ in Eq.(\ref{Psi-def}) describe decoupled right- and left-moving electrons, respectively. In the presence of the electromagnetic field, $\psi_\uparrow$ and $\psi_\downarrow$ modify their  character of  right- and left-movers, although the decoupling in their dynamics is maintained~[\onlinecite{dolcini_2016}]. However, when Rashba interaction is  present, the  dynamics of $\psi_\uparrow$ and $\psi_\downarrow$ is also coupled. This suggests that the  field $\Psi$, i.e. the basis of spin components, is not the most suitable one to obtain the dynamical evolution. In the following subsection, we shall therefore switch to another basis.
 
\subsection{Rotation to the chiral basis}
\label{sec-2B}
We now re-express the electron field spinor $\Psi$ in the spin basis as a rotation around $\sigma_1$ by the space-dependent Rashba angle~$\theta_R(x)$  
\begin{equation}\label{Psi-from-X-compact}
\Psi(x)    \doteq    e^{+\frac{i}{2} \sigma_1\theta_R(x)} \, {\rm X}(x)   
\end{equation}
where
\begin{equation}\label{X-def}
{\rm X}(x)=\left(\begin{array}{l}\chi_{+}(x) \\ \\ \chi_{-}(x) \end{array} \right)
\end{equation}
will be termed the ``chiral'' field spinor, for reasons that will be clear below.  The rotation (\ref{Psi-from-X-compact}), which is inspired by  similar approaches adopted in recent works~[\onlinecite{schmidt_2012},\onlinecite{foster_2016}], 
enables one to rewrite the Hamiltonian $\hat{\mathcal{H}}$~as
\begin{equation}
\hat{\mathcal{H}} =\int dx \,\Psi^\dagger(x) H(x) \Psi^{}(x)     =  \int dx \,{\rm X}^\dagger(x) H_\chi(x) {\rm X}(x) 
\end{equation}
where  
\begin{eqnarray}
 H_\chi(x)  
=  \frac{1}{2}\left\{ v(x)    ,\, {p}_x  -\frac{{\rm e}}{c}  A(x,t) \right\} \sigma_3 \, +{\rm e}   V(x,t) \sigma_0 \hspace{0.5cm} 
\label{Hchi}
\end{eqnarray}
with  
\begin{equation}\label{v(x)-def}
v(x)=\frac{v_F}{\cos\theta_R(x)} = v_F \sqrt{1+\left(\frac{\alpha_R(x)}{\hbar v_F}\right)^2} \, \, \ge v_F 
\end{equation}
The results (\ref{Hchi}) and (\ref{v(x)-def}) deserve a few comments. In the  first instance,  Eq.(\ref{Hchi}) is the first-quantised Hamiltonian for a system of massless Dirac fermions characterised by a {\it space-dependent} velocity  profile (\ref{v(x)-def}), exposed to an electromagnetic field. Secondly, in the chiral basis (\ref{X-def})  such inhomogeneous profile $v(x)$ is the only track left by the Rashba interaction $\alpha_R$, which always {\it increases} the velocity with respect to the bare value of Fermi velocity $v_F$, regardless of the sign of $\alpha_R$.  
Furthermore, in the chiral basis the density (\ref{n-def}) and the current density (\ref{J-def}) acquire  simple expressions, namely
\begin{eqnarray}
\hat{n}(x) &= & \displaystyle  \,{\rm X}^\dagger(x)   \,   {\rm X}(x) =\hat{n}_{+}+\hat{n}_{-} \label{n-chi-fields} \\
\hat{J}(x) &= & \displaystyle v(x) \,{\rm X}^\dagger(x) \sigma_3 \,   {\rm X}(x) =v(x) (\hat{n}_{+}-\hat{n}_{-} ) \label{J-chi-fields} 
\end{eqnarray}
with $\hat{n}_\pm \doteq \chi^\dagger_\pm \chi^{}_\pm$.
Finally, we emphasise that $H_\chi$ is  diagonal  in the chiral basis, implying that the two components~$\chi_\pm$ of  Eq.(\ref{X-def}) are dynamically {\it decoupled}, even when the electromagnetic field is applied and the Rashba interaction is present. 
This feature entails important implications that can be summarized as follows. Let us first consider the case where no electromagnetic field is applied:  in striking contrast with the original spin components $\psi_\uparrow$ and $\psi_\downarrow$, which have a well defined propagation direction only away from the Rashba interaction region, the chiral components $\chi_{+}$ and $\chi_{-}$ describe genuine right-moving and left-moving electrons, respectively, even in the regions where Rashba interaction is present. This explains the origin of the term ``chiral'' and the reason for considering ${\rm X}$ as the ``natural basis'' for Rashba-coupled states. Secondly, when the electromagnetic field is switched on, inelastic and time-reversal breaking processes are induced, and $\chi_\pm$ modify their character of right- and left-movers. Nevertheless, the dynamical decoupling of $\chi_\pm$, encoded in the diagonal structure of Eq.(\ref{Hchi}),  shows that no backscattering arises.
 In the following sections we shall analyze in detail these aspects.

\section{The field-free case: effects of Rashba interaction}
\label{sec-3}
Let us first focus on the effects of the Rashba interaction and switch off the electromagnetic potentials ($A=V=0$). Exploiting the rotation to the chiral basis described in Sec.\ref{sec-2B}, we shall determine the exact evolution $\Psi^\circ(x,t)$  of the electron field for the Hamiltonian $\hat{\mathcal{H}}_\circ$, and then discuss its explicit form for specific examples of Rashba interaction profiles. \\

We start by solving the problem in the chiral basis where the dynamical evolution for the chiral field, dictated by Eq.(\ref{Hchi}), reads
\begin{eqnarray}\label{EOM-Psiprime-fieldfree-pre}
\partial_t {\rm X}^\circ =  -\frac{\sigma_3}{2} \left\{  v(x), \, \partial_x\right\}  {\rm X}^\circ   
\end{eqnarray}
with the superscript ``\,${}^\circ$\,'' in   ${\rm X}^\circ =  (  \chi^\circ_+\,,\, \chi^\circ_- )^T$ reminding that it is the chiral spinor in the absence of electromagnetic  potentials $V$ and $A$. In Eq.(\ref{EOM-Psiprime-fieldfree-pre}) the two components 
$\chi^\circ_+$ and $\chi^\circ_-$
are decoupled, and the solution can be straightforwardly obtained as
\begin{equation}\label{chipm0-sol}
\chi^\circ_\pm(x,t)=  \frac{1}{\sqrt{2 \pi \hbar \,v(x)}} \int dE \, e^{ - i \frac{E}{\hbar}    \left(t \mp \int_{x_{r}}^x \frac{d x^{\prime \prime}}{v(x^{\prime \prime})} \right)}\,\hat{c}_{E\pm}  
\end{equation}
where $\hat{c}^{}_{E+}$ and $\hat{c}^{}_{E-}$ denote fermionic operators for right- and left-moving electrons at the energy $E$, fulfilling $\{ \hat{c}^{}_{E\pm} \, , \hat{c}^\dagger_{E^\prime \pm} \}=\delta(E-E^\prime)$. Furthermore $x_{r}$ is an arbitrarily fixed reference point, such as the space origin or the geometrical center of the Rashba interaction profile.
The exponential of the obtained  solution (\ref{chipm0-sol}) shows that $\chi^\circ_{+}(x,t)$ and $\chi^\circ_{-}(x,t)$ describe genuine right- and left-moving electrons, respectively, propagating with the inhomogeneous velocity (\ref{v(x)-def}).

This is not the case for the dynamical evolution of $\Psi^\circ(x,t)$ in the original spin basis, which is straightforwardly gained by inserting Eq.(\ref{chipm0-sol}) into Eq.(\ref{Psi-from-X-compact}),   obtaining   
\begin{equation}
\psi^\circ_{\uparrow,\downarrow}(x,t)= \int dE \, e^{-i \frac{E t}{\hbar}} \left[\varphi_{E\uparrow,\downarrow}^{(+)}(x) \, \hat{c}_{E+} \,+\, \varphi_{E\uparrow,\downarrow}^{(-)}(x)\, \hat{c}_{E-}  \right] \,.\label{psi-lin-comb}
\end{equation} 
Each spin component at an energy $E$ is thus a combination of   a right-moving $\varphi_{E\uparrow,\downarrow}^{(+)}$ and a left-moving $\varphi_{E\uparrow,\downarrow}^{(-)}$ wave~[\onlinecite{nota-dimensione-phi}], given by
\begin{eqnarray}
\varphi_{E\uparrow}^{(+)}(x) &=& \frac{1}{\sqrt{4 \pi \hbar v(x)}} \,   \sqrt{ 1+\frac{v_F}{v(x)} } \, e^{ + i  s(x)E/\hbar} \, \label{phi-up-RIGHT} \\
\varphi_{E\uparrow}^{(-)}(x) &=& \frac{i \,\mbox{sgn}(\alpha_R(x)) }{\sqrt{4 \pi \hbar v(x) }} \,  \sqrt{ 1-\frac{v_F}{v(x)} }\, e^{ - i  s(x)E/\hbar } \, \label{phi-up-LEFT}  \\
\varphi_{E\downarrow}^{(+)}(x) &=& \frac{i \,\mbox{sgn}(\alpha_R(x)) }{\sqrt{4 \pi \hbar v(x)}} \,   \sqrt{ 1-\frac{v_F}{v(x)} } \, e^{ + i  s(x)E/\hbar } \, \label{phi-dn-RIGHT}  \\
\varphi_{E\downarrow}^{(-)}(x) &=& \frac{1}{\sqrt{4 \pi \hbar v(x) }} \,  \sqrt{ 1+\frac{v_F}{v(x)} }\, e^{ -  i s(x)E /\hbar} \, \label{phi-dn-LEFT}   
\end{eqnarray}
where
\begin{equation}\label{s-def}
s(x) \doteq \int_{x_r}^x \frac{dx^{\prime \prime}}{v(x^{\prime \prime})}
\end{equation}
denotes the ballistic flight-time for the electron to travel from the reference point $x_r$ to $x$, through the spatially varying velocity profile (\ref{v(x)-def}), determined by the Rashba interaction.

\subsection{Model for a Rashba profile}
\begin{figure}[h]
\centering
\includegraphics[width=7cm]{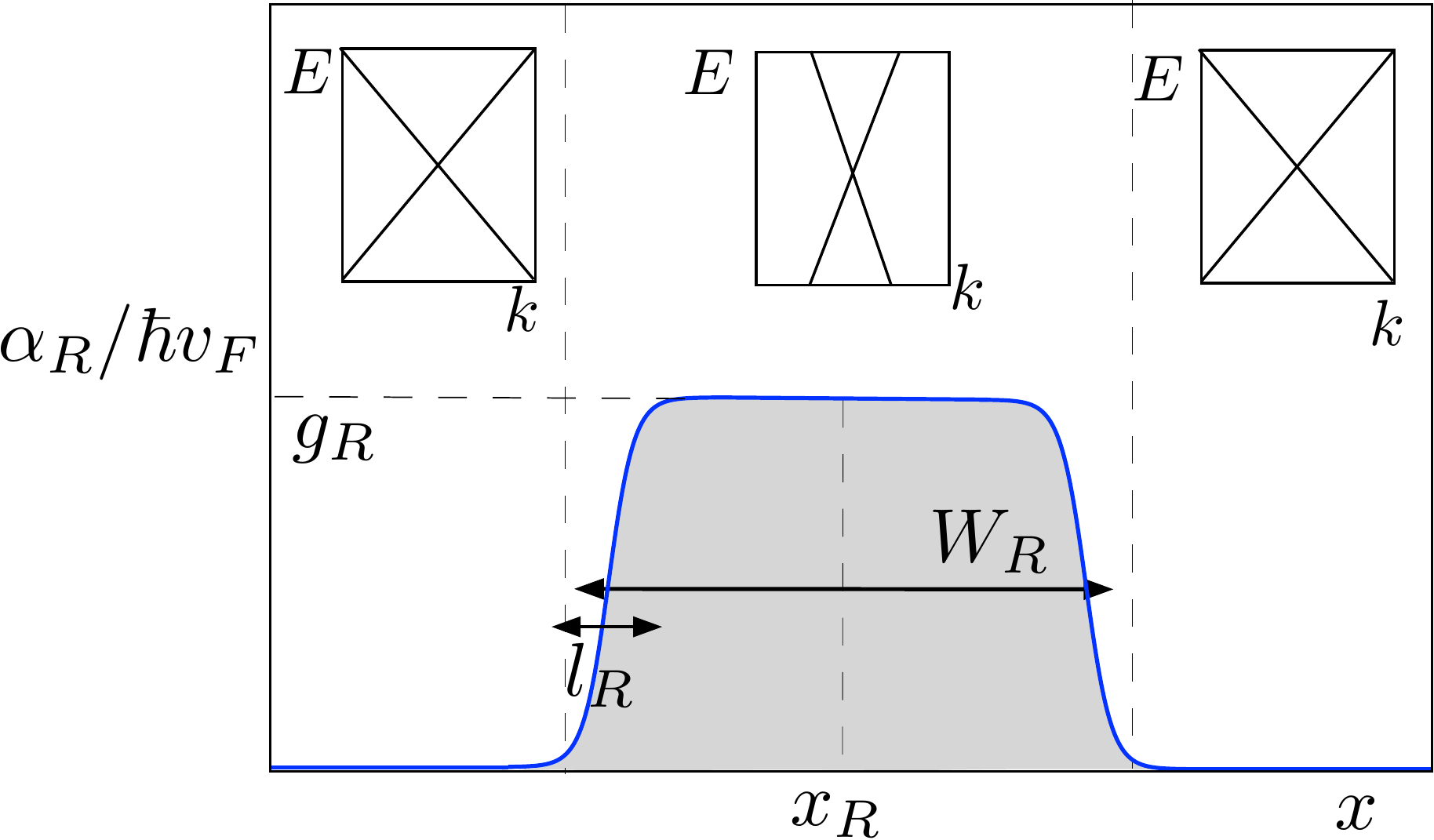}
\caption{
The inhomogeneous profile (\ref{alphaR-region}) modelling a Rashba interaction region. The limit $W_R \ll l_R$ corresponds to a Rashba impurity, while the limit $l_R \ll W_R$ to a Rashba ``barrier''. The insets represent a sketch of the local band-dispersion $E=E(k)$ in the various regions. The Rashba interaction always leads to an increase of the velocity as compared to the bare Fermi velocity $v_F$ of the edge states [see Eq.(\ref{v(x)-def})].
}
\label{Fig-Rashba-profile}
\end{figure}

In order to elucidate the effects of the Rashba interaction, we provide here the explicit behavior of the electron wavefunction for the following model of   Rashba  profile
\begin{equation}\label{alphaR-region}
\alpha_R(x)\, = \hbar \,v_F \, g_R \, \frac{1+e^{-\frac{W_R}{l_R}}}{1+e^{\frac{|x-x_R|-W_R/2}{l_R/2}}}\quad,
\end{equation}
which is centered around the position $x_R$ and characterised by a maximal intensity $g_R$, a width $W_R$ and a smoothening length scale $l_R$, over which it vanishes. The profile, sketched in Fig.~\ref{Fig-Rashba-profile},  can interpolate between two extremal situations: in the limit $W_R \ll l_R$ it describes a sharp Rashba impurity, whose potential extends over a lengthscale $l_R$. In the opposite  limit   $l_R \ll W_R$ it describes a Rashba ``barrier'', where the interaction extends over a scale $W_R$ and abruptly drops  from a roughly constant value to zero~[\onlinecite{nota-barrier}]. This model allows one to obtain an analytical expression of the ballistic flight-time~(\ref{s-def}), which is explicitly given in App.\ref{AppA}.

\begin{figure}[h]
\centering
\includegraphics[width=8cm]{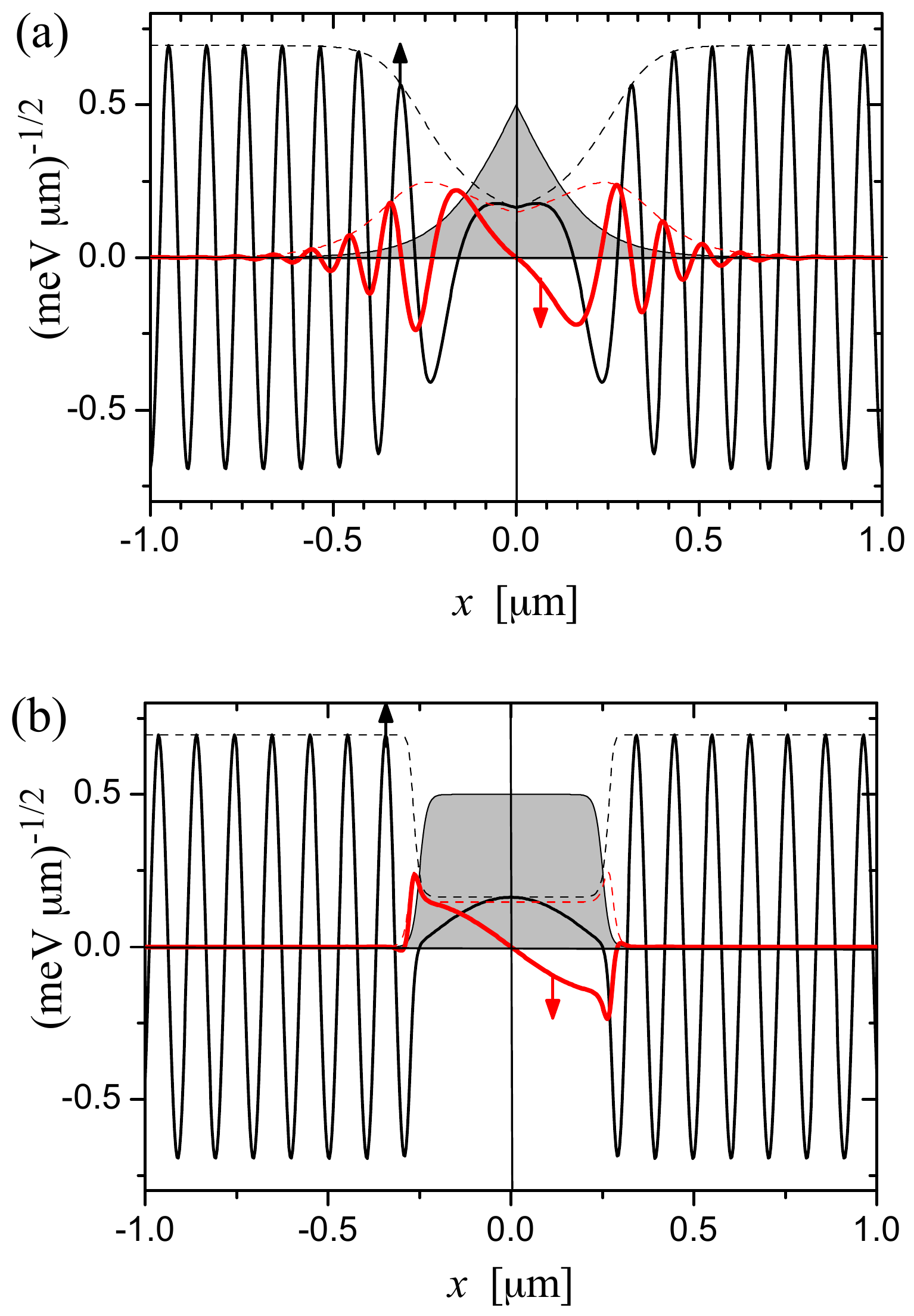}
\caption{
The behavior of the wave function of a right-moving electron approaching a Rashba interaction region, sketched by the grey area in suitably normalised units and modelled by the interaction profile (\ref{alphaR-region}), is shown at the electron energy $E=20\, {\rm meV}$. The real part of  the spin-$\uparrow$ component $\varphi_{E\uparrow}^{(+)}(x)$ [see Eq.(\ref{phi-up-RIGHT})] (thin solid black curve), the spin-$\downarrow$ component $\varphi_{E\downarrow}^{(+)}(x)$ [see Eq.(\ref{phi-dn-RIGHT})] (thick solid red curve) are shown, whereas the dashed curves represent the related absolute values.   Panel (a): the case of a Rashba impurity ($v_F=5 \times  10^5 {\rm m/s}$, $g_R=10$, $x_R=0$ and $l_R=200 \, {\rm nm}$, $W_R=20\,{\rm nm}$). Panel (b): the case of a Rashba barrier ($v_F=5 \times  10^5 {\rm m/s}$,   $g_R=10$, $x_R=0$ and $l_R=20 \, {\rm nm}$, $W_R=500\,{\rm nm}$). While away from the Rashba region a right-moving electron identifies a purely spin-$\uparrow$ component, when the Rashba region is approached a spin-$\downarrow$ component emerges over a length scale dependent on the smoothening length scale $l_R$ of the Rashba profile. Furthermore, the spatial period of the wave function   increases as a result of the velocity enhancement (\ref{v(x)-def})  due to the Rashba interaction. The extremal points of the spin-$\uparrow$ component   correspond to zeros of the spin-$\downarrow$ component.
}
\label{Fig-wave}
\end{figure}

For definiteness, we consider a right-moving electron impinging on the Rashba interaction region at an energy~$E$. In Fig.~\ref{Fig-wave}(a) we show the wave function behavior in the limit $W_R \ll l_R$ of a sharp impurity. The (real part of the) two spin components, $\varphi_{E\uparrow}^{(+)}(x)$ and $\varphi_{E\downarrow}^{(+)}(x)$, are shown by the solid black and red curves, labelled by $\uparrow$ and $\downarrow$, respectively. While away from the Rashba region the right-moving electron is characterised by only spin-$\uparrow$ component, as the Rashba region is approached  an oscillatory spin-$\downarrow$ component arises too. Notice also that the period of the spatial oscillations within the Rashba region is bigger than the one in the bulk of the edge states. Indeed in this region the wavector at a given energy $E$, decreases, as a result of the increase of velocity due to the Rashba interaction (see insets of Fig.~\ref{Fig-Rashba-profile}). 
In Fig.~\ref{Fig-wave}(b) the case  $W_R \gg l_R$ of a Rashba barrier is analyzed. In this case  the spin-$\downarrow$ component arises abruptly as  the Rashba interaction region is entered, and monotonically changes sign across the Rashba region. Notice that in both cases the extremal points of the spin-$\uparrow$ component   correspond to zeros of the spin-$\downarrow$ component.

The imaginary parts of $\varphi_{E\uparrow}^{(+)}(x)$ and $\varphi_{E\downarrow}^{(+)}(x)$, not shown here, behave in a similar way as the real parts shown in Fig.~\ref{Fig-wave}, as can be deduced from the fact that the expressions in Eqs.(\ref{phi-up-RIGHT}) and Eq.(\ref{phi-dn-RIGHT}) differ by a phase $\pm \pi/2$.

\section{Effects of the electromagnetic field}
\label{sec-4}
We now turn to the full problem, switching on the electromagnetic potentials $V,A$. Here below we  derive the exact dynamical evolution of the electron field. 
\subsection{Solution in the chiral basis}
Again, it is first worth solving the equation  of motion in the chiral basis ${\rm X}$, obtaining from Eq.(\ref{Hchi})   two   decoupled equations
\begin{eqnarray}
\lefteqn{
 \left(\partial_t \pm v(x)\partial_x \pm\frac{1}{2} \partial_x v(x)   \right) \chi_\pm =  } & & \nonumber \\
 & &  \, - \frac{i{\rm e}}{\hbar} \left(    V(x,t)    \mp  \frac{v(x)}{c}\,A(x,t) \, \right)  \chi_\pm \quad. \label{eq-chi-with-field} 
\end{eqnarray}
It is possible to verify that the solution of Eq.(\ref{eq-chi-with-field}) is
\begin{equation}\label{chipm-sol}
\chi^{}_\pm(x,t) =   e^{\pm i \phi_\pm(x,t)}\, \chi^\circ_\pm(x,t) 
\end{equation}
where $\chi^\circ_\pm(x,t)$ is a solution (\ref{chipm0-sol}) for the field-free case, while

\begin{equation}\label{phipm}
\begin{array}{l}
\phi_+(x,t) = \displaystyle {\frac{{\rm e}}{\hbar c} \int_{-\infty}^x \!\!\!\! dx^\prime  \left( A-\frac{c}{v(x^\prime)}V\right) (x^\prime,t-\int_{x^\prime}^{x} \frac{dx^{\prime \prime}}{v(x^{\prime\prime})}) }   \\  \\
\phi_-(x,t) = \displaystyle  {\frac{{\rm e}}{\hbar c} \int_{x}^{\infty}  \!\!\!\! dx^\prime   \left( A+\frac{c}{v(x^\prime)}V\right) (x^\prime,t-\int_{x}^{x^\prime} \frac{dx^{\prime \prime}}{v(x^{\prime\prime})}) }   
\end{array}  
\end{equation}
describe the phases induced by the electromagnetic field. As one can see, the phase $\phi_+$ ($\phi_-$) at a given space-time point $(x,t)$ is expressed as a  space  convolution  of the electromagnetic potentials $A$ and $V$, evaluated at an earlier time that depends on the inhomogeneous velocity profile  (\ref{v(x)-def}) dictated by the Rashba interaction. The result (\ref{chipm-sol}) thus fully describes --within the assumption of independent electrons-- the electron dynamics in the presence of Rashba interaction and electromagnetic field.  \\

In the chiral basis, the gauge invariant correlation function is defined as
\begin{eqnarray}
-i G^{<}_{\chi \pm}(x,t;x^\prime,t^\prime) &\doteq & \langle \chi^\dagger_{\pm} (x^\prime,t^\prime)\chi_\pm(x,t)\rangle_\circ \, \times  \nonumber \\
& & \times \, e^{-i\frac{\rm e}{\hbar c} \int_{(x,t)}^{(x^\prime,t^\prime)} (c V d  t^{\prime\prime}-A dx^{\prime\prime})}.\hspace{1cm} \label{Gchi-def}
\end{eqnarray}
Here the expectation value $\langle \ldots \rangle_\circ$ is computed with respect to the  equilibrium state at   $t=-\infty$, where the electromagnetic potentials are assumed to be switched off and the dynamics is dictated by $\hat{\mathcal{H}}_\circ$ containing the linear band term and the Rashba interaction.
Notably, the phase factor in the second line of Eq.(\ref{Gchi-def}) is the Wilson line that ensures the gauge invariance of the correlation~[\onlinecite{bertlmann}].
Equation (\ref{Gchi-def}) is thus straightforwardly evaluated by exploiting the solution Eq.(\ref{chipm-sol}) and the   equilibrium correlation function,   
\begin{eqnarray}
\langle {\chi^\circ_\pm}^\dagger (x^\prime,t^\prime) {\chi^\circ_\pm}^{}(x,t)  \rangle_\circ = -i \, \frac{e^{\frac{i}{\hbar} E_F (t^\prime-t\mp \int_{x}^{x^\prime} \frac{d x^{\prime \prime}}{v(x^{\prime \prime})})}}{\beta \hbar \sqrt{v(x) \, v(x^\prime)}} \times \nonumber \\
 \displaystyle  \,\times  \frac{1}{2 \sinh\left[ \pi   ( 
t^\prime-t\mp \int_{x}^{x^\prime} \frac{d x^{\prime \prime}}{v(x^{\prime \prime})})/\beta \hbar \right]} \hspace{1cm} \label{corr-0}
\end{eqnarray}
of the chiral fields, where $\beta=\/k_B T$ is the inverse temperature and $E_F$ the Fermi energy.\\

In particular, the {\it photoexcited} chiral density $\Delta {n}_{\pm}$, which physically describes the deviation in the expectation value of $\hat{n}_\pm=\chi^\dagger_\pm \chi^{}_\pm$  from its equilibrium value, is 
mathematically defined from Eq.(\ref{Gchi-def}) by subtracting its value at vanishing electromagnetic field and by taking the limit $(x^\prime,t^\prime) \rightarrow (x,t)$, i.e.
\begin{eqnarray}
\Delta  {n}_\pm(x,t)   \doteq -i \!\!\!\!\lim_{\epsilon_x,\epsilon_t \rightarrow 0} \! \!\left[ G^{<}_{\chi \pm}(x-\frac{\epsilon_x}{2},t-\frac{\epsilon_t}{2};x+\frac{\epsilon_x}{2},t+\frac{\epsilon_t}{2})\right.    \nonumber \\
 \left. \left. -G^{<}_{\chi \pm}(x-\frac{\epsilon_x}{2},t-\frac{\epsilon_t}{2};x+\frac{\epsilon_x}{2},t+\frac{\epsilon_t}{2})\right|_{A,V=0}\right]\, .\hspace{1cm} 
\end{eqnarray}
Proceeding in a similar way as it was done in Ref.[\onlinecite{dolcini_2016}], one obtains  three equivalent  expressions for $\Delta  {n}_\pm$, namely
\begin{eqnarray}
\Delta  {n}_+(x,t) &=& \displaystyle  +\frac{{\rm e}}{2\pi\hbar v(x)} \int_{-\infty}^x dx^\prime E(x^\prime,t-\int_{x^\prime}^{x} \frac{dx^{\prime \prime}}{v(x^{\prime \prime})}) =\nonumber  \\  
&=& \displaystyle \frac{1}{2\pi}   \,  \left( \partial_x  \phi_{+} - \frac{{\rm e}}{\hbar c}  A \right) =  \label{Delta-n_+}\\  
&=&  \displaystyle    \frac{1}{2\pi v(x)}   \,  \left( - \partial_t  \phi_{+}  - \frac{{\rm e}}{\hbar} V \right) \nonumber
\end{eqnarray}
and
\begin{eqnarray}
\Delta  {n}_-(x,t)
&=& \displaystyle-\frac{{\rm e}}{2\pi\hbar v(x)} \int_{x}^{\infty}  dx^\prime E(x^\prime,t-\int_{x}^{x^\prime} \frac{dx^{\prime \prime}}{v(x^{\prime \prime})}) =\nonumber \\  
&=& \displaystyle \frac{1}{2\pi}   \,  \left( \partial_x  \phi_{-} + \frac{{\rm e}}{\hbar c}  A \right) = \label{Delta-n_-} \\  
&=&  \displaystyle    \frac{1}{2\pi v(x)}   \,  \left( +\partial_t  \phi_{-}  - \frac{{\rm e}}{\hbar} V \right)  \nonumber \quad.
\end{eqnarray}
The expressions obtained in first lines of Eqs.(\ref{Delta-n_+})-(\ref{Delta-n_-}) are particularly insightful.  In the first instance they are gauge invariant, as it should be, since they   purely depend on the electric field $E(x,t)$ and not on the specific choice of the potentials $V$ and $A$. The inclusion of the Wilson line in Eq.(\ref{Gchi-def}) is crucial to this purpose. Second, $\Delta n_\pm$ are expressed as convolutions of the electric field evaluated at earlier times, where the Rashba interaction enters as a retardation effect via the inhomogeneous velocity  (\ref{v(x)-def}). This is similar to the phases in Eqs.(\ref{phipm}) that are, however, gauge-dependent. Third, the time-dependence of Eqs.(\ref{Delta-n_+}) and (\ref{Delta-n_-}) shows that $\Delta n_\pm$ are first gradually ``created'' by the electric pulse in the area where $E(x,t)$ is applied, while only at locations $x$ far away from such area, where the upper/lower boundary of the integral can effectively be sent to $\pm\infty$, the profiles $\Delta n_\pm$ evolve as right- and left-moving packets. Their profile can be modified by the Rashba interaction via the velocity,  as  will be shown in Sec.\ref{sec-5} by explicit examples.
Finally, it is worth stressing that the obtained $\Delta n_\pm$ are independent of the temperature and of the chemical potential of the initial equilibrium state of the edge states with Rashba  interaction. This generalizes the results of Ref.[\onlinecite{dolcini_2016}] to the presence of Rashba interaction.  
\subsection{Chiral anomaly in the presence of Rashba interaction}
Taking time and space derivatives of Eqs.(\ref{Delta-n_+}) and (\ref{Delta-n_-}),  one obtains
\begin{eqnarray}
\partial_t \Delta \hat{n}_{+} +\partial_x [v(x) \Delta \hat{n}_{+}]=+\frac{{\rm e}}{2\pi \hbar} E(x,t) \label{chi-ano-1-a}\, \,\\
\partial_t \Delta \hat{n}_{-} -\partial_x [v(x) \Delta \hat{n}_{-}]=-\frac{{\rm e}}{2\pi \hbar} E(x,t) \label{chi-ano-1-b}\,,
\end{eqnarray}
The terms appearing on the right-hand side of Eqs.(\ref{chi-ano-1-a}) and (\ref{chi-ano-1-b}) break the conservation laws that one would obtain from Eqs.(\ref{eq-chi-with-field}) by treating the fields $\chi_\pm$ classically.
This generalises the chiral anomaly effect~[\onlinecite{adler1969,bell-jackiw1969,nielsen1983}], i.e. the  response of massless Dirac fermions to the electromagnetic field, to the case of inhomogeneous velocity profile.
Equivalently, by taking sum and difference of Eqs.(\ref{chi-ano-1-a}) and (\ref{chi-ano-1-b}) and by recalling Eqs.(\ref{n-chi-fields})-(\ref{J-chi-fields}) and (\ref{n-def})-(\ref{J-def}), one obtains
\begin{eqnarray}
\partial_t \Delta \hat{n}  +\partial_x \Delta \hat{J} &=&0 \label{chi-ano-2-a}\\
\partial_t \Delta \hat{n}^a -\partial_x \Delta \hat{J}^a &=& \frac{{\rm e}}{\pi \hbar} E(x,t)  \quad, \label{chi-ano-2-b}
\end{eqnarray}
where Eq.(\ref{chi-ano-2-a}) is the continuity equation describing the conservation of electrical charge, while Eq.(\ref{chi-ano-2-b})  describes the  anomalous breaking of the conservation law that  involve the quantities 
\begin{eqnarray} 
\hat{n}^a &=&\Psi^\dagger (\cos\theta_R \sigma_3+\sin\theta_R \sigma_2) \Psi^{}={\rm X}^\dagger \sigma_3 {\rm X}    \label{axial-charge} \\
\hat{J}^a &=& \frac{v_F}{\cos\theta_R(x)} \Psi^\dagger \Psi^{}=v(x) \,{\rm X}^\dagger {\rm X} \,. \label{axial-current}   
\end{eqnarray} 
Equations (\ref{axial-charge}) and (\ref{axial-current}) extend the definition of the axial charge and current densities to the case where Rashba interaction is present. Note that the anomalous term on the right-hand side of Eqs.(\ref{chi-ano-1-a}), (\ref{chi-ano-1-b}) and (\ref{chi-ano-2-b}) purely depends on the electric field and the universal constant ${\rm e}/\pi \hbar$, and is independent of the electronic state.
\subsection{Solution in the spin basis}
The dynamical evolution in the original spin basis $\Psi$ can be found by applying the transformation (\ref{Psi-from-X-compact}), i.e.
\begin{equation}\label{Psiprime-Psi}
\left(\begin{array}{l}\psi_{\uparrow}(x) \\ \\ \psi_{\downarrow}(x) \end{array} \right)  = \left( 
\begin{array}{cc} \cos\frac{\theta_R}{2} & i \sin\frac{\theta_R}{2} \\ & \\i \sin\frac{\theta_R}{2} & \cos\frac{\theta_R}{2} \end{array} \right) \left(\begin{array}{l}\chi_{+}(x) \\ \\ \chi_{-}(x) 
\end{array} \right) \, ,   
\end{equation}
to the solution (\ref{chipm-sol}) obtained in the chiral basis ${\rm X}$. The correlation functions in the spin basis are then also straightforwardly obtained from Eqs.(\ref{Gchi-def}) or (\ref{Delta-n_+})-(\ref{Delta-n_-}) via the rotation (\ref{Psiprime-Psi}). In particular, one obtains for the photoexcited local spin densities and correlations  
\begin{widetext}{\small 
\begin{eqnarray}
\Delta  {n}_{\uparrow}(x,t) &=&   \frac{1}{2\pi\hbar v(x)}\left[ \frac{1+\frac{v_F}{v(x)}}{2}     \int_{-\infty}^x dx^\prime E\left(x^\prime,t-\int_{x^\prime}^{x} \frac{dx^{\prime \prime}}{v(x^{\prime \prime})}\right)   -\frac{1-\frac{v_F}{v(x)}}{2}  \int_{x}^{\infty}  dx^\prime E\left(x^\prime,t-\int_{x}^{x^\prime} \frac{dx^{\prime \prime}}{v(x^{\prime \prime})}\right) \right] \hspace{1cm} \label{Delta-n_up}\\
\Delta  {n}_{\downarrow}(x,t) &=&   \frac{1}{2\pi\hbar v(x)} \left[\frac{1-\frac{v_F}{v(x)}}{2} \int_{-\infty}^x dx^\prime E\left(x^\prime,t-\int_{x^\prime}^{x} \frac{dx^{\prime \prime}}{v(x^{\prime \prime})}\right) -\frac{1+\frac{v_F}{v(x)}}{2}  \int_{x}^{\infty}  dx^\prime E\left(x^\prime,t-\int_{x}^{x^\prime} \frac{dx^{\prime \prime}}{v(x^{\prime \prime})}\right)   \right]\hspace{1cm} \label{Delta-n_dn} \\
\Delta \langle\psi^\dagger_{\uparrow}\psi^{}_{\downarrow}\rangle(x,t) &=&  + \frac{i \,\sin(\alpha_R(x))}{4\pi\hbar v(x)}\,  \left[  \int_{-\infty}^x dx^\prime E\left(x^\prime,t-\int_{x^\prime}^{x} \frac{dx^{\prime \prime}}{v(x^{\prime \prime})}\right) + \int_{x}^{\infty}  dx^\prime E\left(x^\prime,t-\int_{x}^{x^\prime} \frac{dx^{\prime \prime}}{v(x^{\prime \prime})}\right)   \right] \label{Delta-psi_up-psi_dn} \hspace{1cm}\\
\Delta \langle\psi^\dagger_{\downarrow}\psi^{}_{\uparrow}\rangle(x,t) &=&   -\frac{i \,\sin(\alpha_R(x))}{4\pi\hbar v(x)}\,      \left[ \int_{-\infty}^x dx^\prime E\left(x^\prime,t-\int_{x^\prime}^{x} \frac{dx^{\prime \prime}}{v(x^{\prime \prime})}\right) + \int_{x}^{\infty}  dx^\prime E\left(x^\prime,t-\int_{x}^{x^\prime} \frac{dx^{\prime \prime}}{v(x^{\prime \prime})}  \right) \right] \label{Delta-psi_dn-psi_up}  \hspace{1cm}
\end{eqnarray}
}
\end{widetext}
A few comments about the results (\ref{Delta-n_up}) to (\ref{Delta-psi_dn-psi_up}) are in order. In the first instance, the photoexcited quantities only depend on the applied electric field and are {\it independent} of the temperature and the chemical potential of the initial equilibrium state, extending the results of Ref.[\onlinecite{dolcini_2016}] to the case where  Rashba interaction is present. 
Secondly,  the photoexcited spin-$\uparrow$ electron density $\Delta  {n}_{\uparrow}$, Eq.(\ref{Delta-n_up}), consists of two terms, which can be well identified asymptotically, i.e. for  for positions $x$ away from the spatial extension of the electromagnetic pulse: the first term describes a right-moving component, while the second one --which vanishes in the absence of a Rashba coupling-- describes a left-moving component. A similar remark holds for $\Delta  {n}_{\downarrow}$.  Furthermore, the photoexcitation also leads to the appearance of expectation values $\langle\psi^\dagger_{\uparrow}\psi^{}_{\downarrow}\rangle$ and $\langle\psi^\dagger_{\downarrow}\psi^{}_{\uparrow}\rangle$, which are vanishing in the absence of Rashba coupling. We emphasise that this {\it does not}
correspond to any electron back-scattering, which would be encoded in a non-vanishing mixed transition amplitude $\langle\chi^\dagger_{+}\chi^{}_{-}\rangle$ of the right- and left-moving chiral fields, induced by the electromagnetic field. Here, as observed above, backscattering is absent since the electromagnetic field does not couple the $``+''$ and $``-''$ chiral sectors [see Eqs.(\ref{eq-chi-with-field})],   despite  breaking  time-reversal symmetry and inducing inelastic processes. The non vanishing values of $\Delta \langle\psi^\dagger_{\uparrow}\psi^{}_{\downarrow}\rangle$ is thus  a mere consequence of the fact since spin-$\uparrow$ and spin-$\downarrow$ components do not correspond to eigenstates when Rashba interaction is present. 

Note that, in particular, the {\it total} photoexcited density acquires a compact expression
\begin{eqnarray}
\lefteqn{\Delta  {n}(x,t) =\Delta  {n}_{\uparrow}(x,t)+\Delta {n}_{\downarrow}(x,t) \, \,= } & & \nonumber \\
&=&\, \, \Delta {n}_{+}(x,t)+\Delta  {n}_{-}(x,t)\, =  \nonumber \\
&=& \, \frac{1}{2\pi\hbar v(x)}\left[ \int_{-\infty}^x dx^\prime E\left(x^\prime,t-\int_{x^\prime}^{x} \frac{dx^{\prime \prime}}{v(x^{\prime \prime})}\right)  -  \right.\nonumber \\
& & \hspace{1.3cm} -\left.\int_{x}^{\infty}  dx^\prime E\left(x^\prime,t-\int_{x}^{x^\prime} \frac{dx^{\prime \prime}}{v(x^{\prime \prime})}\right)  \right] \label{dens-tot}\hspace{1cm} 
\end{eqnarray}
where the two terms on the right-hand of the last line asymptotically correspond to the right-moving and a left-moving contributions, respectively.

\section{Gaussian electric pulse in the presence of a Rashba barrier}
\label{sec-5}
In this section we shall apply the general results obtained in the previous sections to the case of a Gaussian electric pulse,  directed along the propagation direction of the edge states and applied for a duration $\tau$ over a finite region of size $\Delta$. We shall choose the origin of  $x$-axis in the spatial center of the pulse,   
\begin{equation}\label{E-gauss-pulse}
E(x,t)= E_0 \, e^{-\frac{x^2}{2 \Delta^2}} \, \, e^{-\frac{t^2}{2 \tau^2}}  \, 
\end{equation}
and we shall assume that a Rashba interaction region, centered around the position $x_R$ and extending over a width $W_R$, exists along such direction  and can be described by the profile model (\ref{alphaR-region}).  
The total photoexcited electron density  (\ref{dens-tot}) consists of the two terms Eqs.(\ref{Delta-n_+}) and (\ref{Delta-n_-}) that, in the case of  the Gaussian pulse (\ref{E-gauss-pulse}), reduce to
\begin{eqnarray}
\Delta  {n}_+(x,t) = \displaystyle  +\frac{{\rm e} E_0}{2\pi\hbar} \frac{1}{v(x)}\int_{-\infty}^x \hspace{-0.3cm} dx^\prime e^{-\frac{{x^\prime}^2}{2 \Delta^2}} \,e^{-\frac{(s(x^\prime)-s(x)+t)^2}{2 \tau^2}} \hspace{0.3cm} \label{Delta-n_{+}-Gauss-pre}\\
\Delta  {n}_-(x,t) = \displaystyle  -\frac{{\rm e} E_0}{2\pi\hbar} \frac{1}{v(x)}\int_{x}^{\infty} \hspace{-0.3cm}dx^\prime e^{-\frac{{x^\prime}^2}{2 \Delta^2}} \,e^{-\frac{(s(x^\prime)-s(x)-t)^2}{2 \tau^2}} \hspace{0.3cm}\label{Delta-n_{-}-Gauss-pre}
\end{eqnarray}
where $s(x)$ is the ballistic flight-time (\ref{s-def}) determined by the Rashba profile,  given by Eq.(\ref{s-model}) for the model (\ref{alphaR-region}).
For definiteness, we shall analyze here the limit of ``Rashba barrier'',  $\lambda_R \ll W_R$ , where $s(x)$ acquires the simplified expression (\ref{s-model-box}). The photoexcited density components (\ref{Delta-n_{+}-Gauss-pre}) and (\ref{Delta-n_{-}-Gauss-pre}) can be straightforwardly computed by numerical integration. Here below we discuss various situations.
\subsection{Photoexcitation far from the Rashba region}
\label{sec-5-a}
\begin{figure}[h]
\centering
\includegraphics[width=\columnwidth]{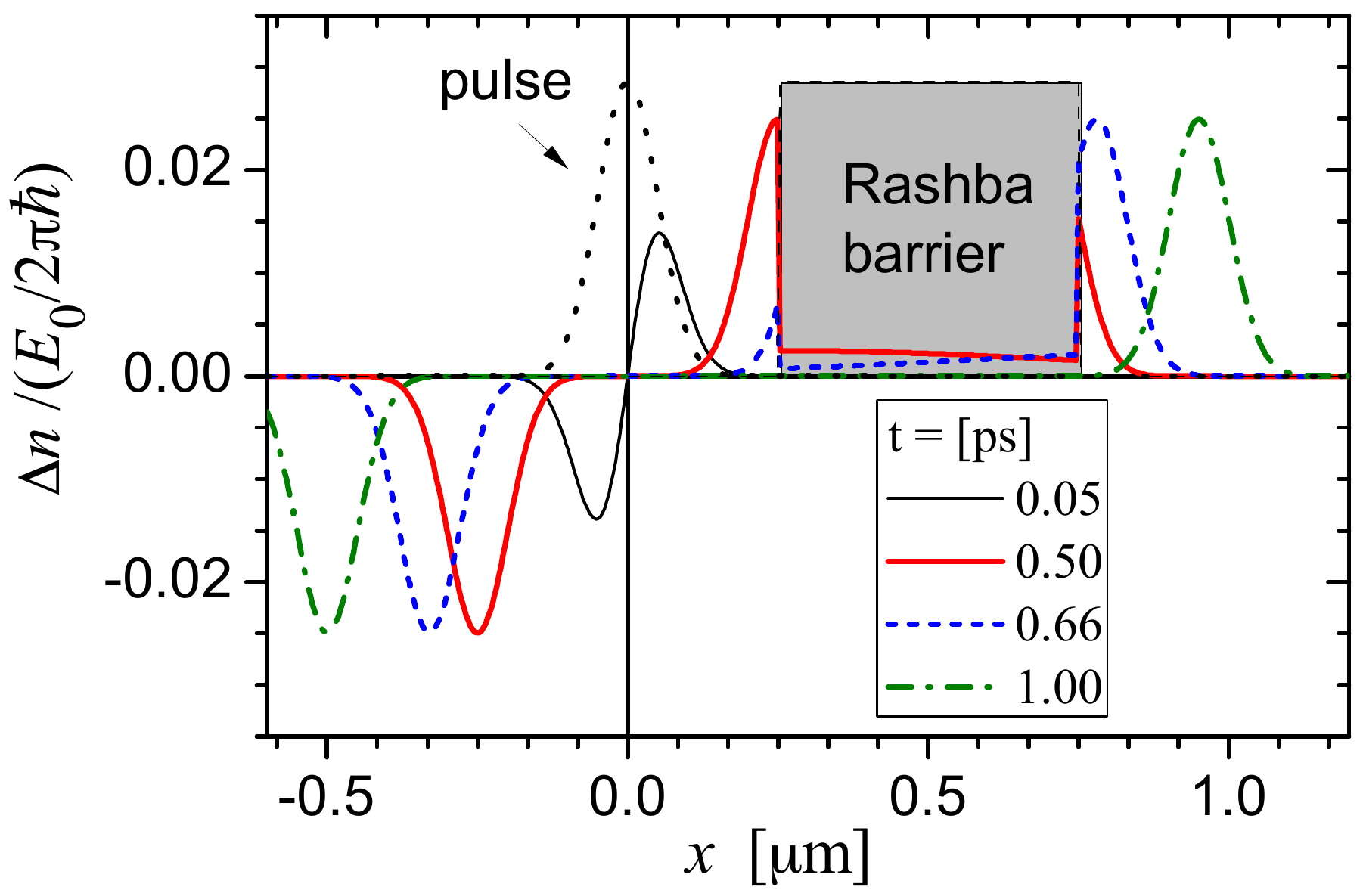}
\caption{
The case of photoexcitation occurring far from the Rashba region. A Gaussian electric pulse (\ref{E-gauss-pulse}) with $\Delta=50\,{\rm nm}$ and $\tau=10\,{\rm fs}$ (spatial profile sketched by the dotted curve in suitable units) is applied far away from a Rashba interaction region (sketched by the grey area in suitable units) characterized by parameters $W_R=500\,{\rm nm}$, $x_R=500 \,{\rm nm}$, $g_R=10$ and $l_R \rightarrow 0$ in (\ref{alphaR-region}). The photoexcited  electron density $\Delta n$ [see Eq.(\ref{dens-tot})]  is shown at various snapshots: $t=0.05 \,{\rm ps}$ (thin solid curve), $t=0.5 \,{\rm ps}$ (solid curve), $t=0.66 \,{\rm ps}$ (dashed curve) and $t=1.0 \,{\rm ps}$ (dash-dotted curve). Due to the increase of velocity (\ref{v(x)-def}) ascribed to the Rashba interaction, the  right-moving wavepacket is ``boosted'' by the Rashba barrier, as compared to its left-moving photoexcited partner that does not impact the barrier.  The Rashba region effectively acts  as a ``superluminal gate'' that enables the wave packet to access space-time regions beyond the light cone characterised by the bare Fermi velocity $v_F$.
}
\label{Fig-dens-1}
\end{figure}

The first situation  we analyze is the case where the Gaussian electric pulse (\ref{E-gauss-pulse}) and the Rashba region (\ref{alphaR-region})  are spatially separated, so that they do not direct interplay. This is the case illustrated in Fig.~\ref{Fig-dens-1}, where the dotted curve and the grey area show, in suitably normalized units, the electric pulse profile and the Rashba interaction profile, respectively. The former is centered at the origin and characterised by a lengthscale $\Delta=50\,{\rm nm}$, while the latter is centered at $x_R=500 \,{\rm nm}$ over a lengthscale $W_R=500\,{\rm nm}$. 
The Gaussian pulse photoexcites an electron density $\Delta n$ that consists of two counterpropagating wavepackets of opposite sign (no net charge is created), with the right-moving one reaching the Rashba region. Here, due to the increase of velocity (\ref{v(x)-def}) caused by the Rashba interaction, the front of the  electron wavepacket ``slips'' across the Rashba region before its tail has entered it.  
At long timescales, the space profile of the wavepacket eventually emerging from the Rashba profile  is unaffected by it, and is purely determined by the parameters $\Delta$ and $\tau$ of the initially applied pulse.  In particular, its asymptotic shape is Gaussian, with a standard deviation $D=\sqrt{\Delta^2+(v_F \tau)^2}$. \\ Thus, in this case the Rashba interaction effectively acts as a ``superluminal gate'' that shuttles the wavepacket to space-time regions located outside the light-cone of the bare Fermi velocity $v_F$. This clearly appears when comparing the locations of the two counterpropagating partner wavepackets at $t=1\, {\rm ps}$ (dash-dotted green curves): while in the absence of Rashba interaction they would be located at symmetric positions $x=\pm 0.5 \mu{\rm m}$, the right-moving one has been boosted by the Rahsba region further off from the origin, as compared to the left-moving one.

\subsection{Photoexcitation inside the Rashba region}
\label{sec-5-b}
\begin{figure}[h]
\centering
\includegraphics[width=\columnwidth]{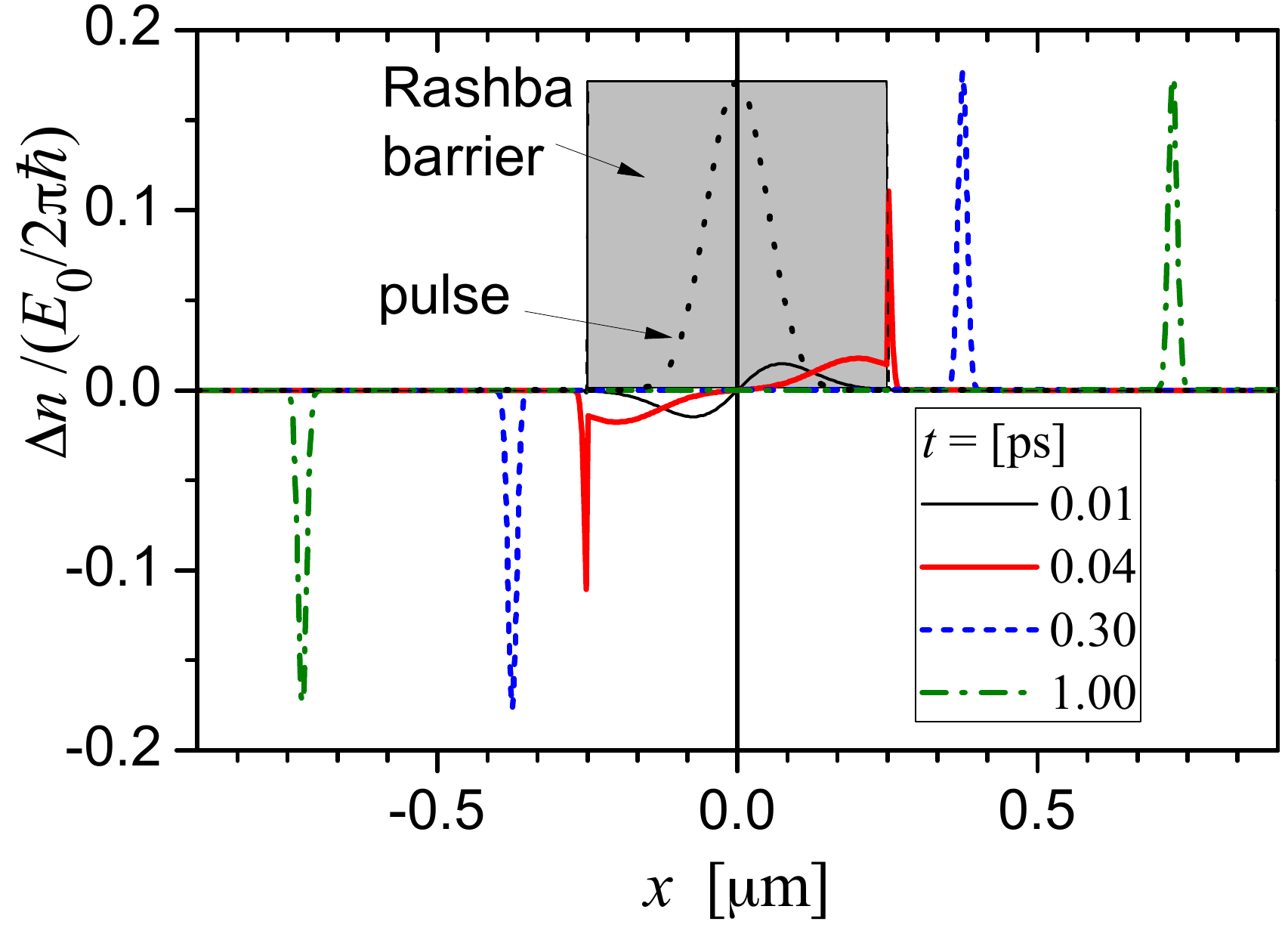}
\caption{
The case of electron photoexcitation occurring inside the Rashba region. A Gaussian electric pulse (\ref{E-gauss-pulse}) with $\Delta=50\,{\rm nm}$ and $\tau=10\,{\rm fs}$ (spatial profile sketched by the dotted curve in suitable units) is applied inside a Rashba barrier with parameters $W_R=500\,{\rm nm}$, $x_R=0$, $g_R=10$ and $l_R \rightarrow 0$ in (\ref{alphaR-region}) (grey area, normalised units). The photoexcited  electron density $\Delta n$ [see Eq.(\ref{dens-tot})]  is shown at various snapshots: $t=0.01 \,{\rm ps}$ (thin solid curve), $t=0.04 \,{\rm ps}$ (solid curve), $t=0.3 \,{\rm ps}$ (dashed curve) and $t=1.0 \,{\rm ps}$ (dash-dotted curve). When the wavepacket emerges from the Rashba interaction region where photoexcitation occurs, it experiences a decrease of velocity from the value (\ref{v(x)-def}) to the bare Fermi velocity $v_F$, and its profile gets squeezed.
}
\label{Fig-dens-2}
\end{figure}

The second situation we analyze is the case where the photoexcitation occurs fully inside the Rashba region   and directly interplays with it.  
We consider for instance the same Gaussian electric pulse as in Fig.~\ref{Fig-dens-1},  now applied at the center of the Rashba region, as illustrated by dotted curve and the grey area  of Fig.~\ref{Fig-dens-2}, respectively,  in suitably normalized units.   Again the pulse photoexcites two counterpropagating wavepackets of opposite sign, which  evolve symmetrically in this case. Note that, since the Gaussian pulse extends over a shorter lengthscale than the Rashba region ($3\Delta<W_R$), the photoexcitation occurs fully inside the Rashba region, where the velocity is higher than the bare Fermi velocity [see Eq.(\ref{v(x)-def})]. Thus, when the photoexcited wavepackets emerge from both sides of the Rashba region, the velocity ``slows down'' to the value $v_F$, and a squeezing of the spatial profile occurs, as can be seen in the curves at $t=0.3\, {\rm ps}$ and $t=1.0\,{\rm ps}$ in  Fig.~\ref{Fig-dens-2}. Note the corresponding enhancement of the vertical scale in Fig.~\ref{Fig-dens-2} as compared to  Fig.~\ref{Fig-dens-1}. 
This squeezing effect can be understood by the following intuitive argument: an electric pulse with amplitude $E_0$ applied for a time $\tau$  photoexcites electrons within a certain energy scale $\Delta E$, roughly determined by $v \tau {\rm e}E_0$, where $v$ is the electron group velocity in the Rashba barrier, where the pulse is applied. In turn, $\Delta E$ identifies a range $\Delta k$ of electronic states in the band dispersion of the Rashba region (see central inset of Fig.~\ref{Fig-Rashba-profile}). After the ending of the pulse, the electron dynamics is elastic, so that   the same energy scale $\Delta E$ identifies a larger range $\Delta k^\prime$ of electronic   states when the group velocity is reduced down to $v_F$ outside the Rashba barrier (side insets of Fig.~\ref{Fig-Rashba-profile}). By Heisenberg uncertainty principles, the wavepacket spatial extension then shrinks roughly by an amount $\sqrt{1+g_R^2}$. This argument qualitatively explains the gist of the effect, although it is  quantitatively not rigorous, since the spatial scale $\Delta$ of the applied pulse plays a role too.\\
Note that, in striking contrast with the situation described in Sec.\ref{sec-5-a}, here the Rahsba interaction directly affects the final shape of the propagating wavepackets.

\subsection{Photoexcitation  partially overlapping with the Rashba region}
\label{sec-5-c}
\begin{figure}[h]
\centering
\includegraphics[width=\columnwidth]{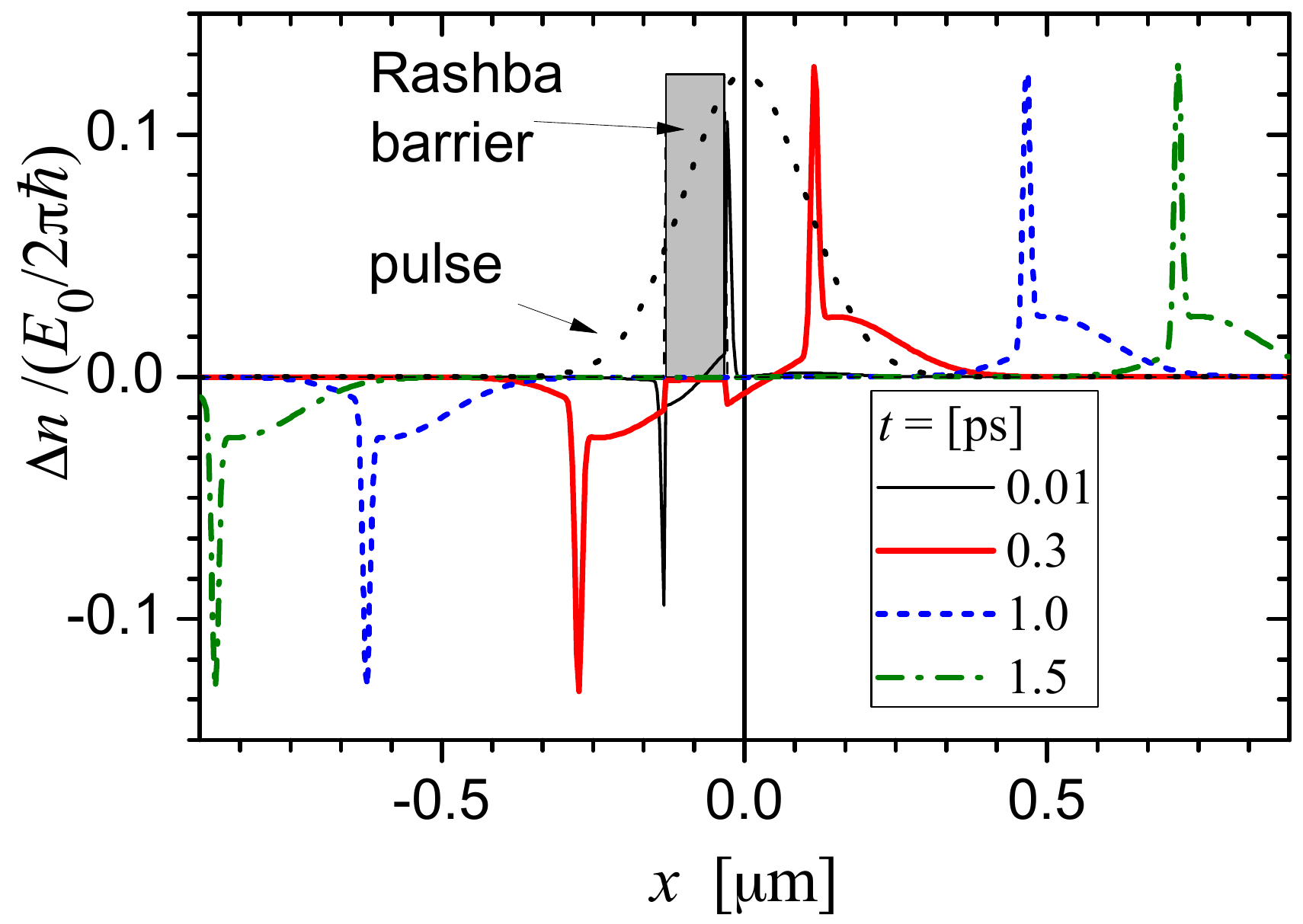}
\caption{
The case of photoexcitation  partially overlapping with the Rashba region. A Gaussian electric pulse (\ref{E-gauss-pulse}) with $\Delta=100\,{\rm nm}$ and $\tau=10\,{\rm fs}$ (spatial profile sketched by the dotted curve in suitable units) is applied in the presence of a Rashba interaction barrier with parameters $W_R=100\,{\rm nm}$, $x_R=-80 \,{\rm nm}$, $g_R=10$ and $l_R \rightarrow 0$ in (\ref{alphaR-region}) (grey area, normalised units). The photoexcited  electron density $\Delta n$ [see Eq.(\ref{dens-tot})]  is shown at various snapshots: $t=0.01 \,{\rm ps}$ (thin solid curve), $t=0.3 \,{\rm ps}$ (solid curve), $t=1.0 \,{\rm ps}$ (dashed curve) and $t=1.5 \,{\rm ps}$ (dash-dotted curve). The wavepackets emerging from the Rashba  region exhibit  a fuzzy profile where two different Gaussian-like profiles coexist. The sharp peak stems from the photoexcitation occurring inside the Rashba barrier, whereas the broader one originates from the photoexcitation processes outside the Rashba barrier.
}
\label{Fig-dens-3}
\end{figure}

The  third interesting situation is when there is a partial overlap between the photoexcitation area and the Rashba region. Consider now the Gaussian electromagnetic  pulse illustrated --again, in suitably normalized units-- by the dotted curve  of Fig.~\ref{Fig-dens-3}, which is applied over a  lengthscale   extending beyond the Rashba region ($3\Delta>W_R$) depicted by the grey area. In this case the photoexcitation process takes place both inside and outside the Rashba region, and two different velocities are involved in it. As a consequence, the photoexcited wavepackets acquire a fuzzy shape characterized by a broader Gaussian profile, arising from the photoexcitation outside the Rashba barrier, and a narrower Gaussian peak ascribed to the states originally photoexcited inside the Rashba barrier.  The asymptotically propagating wave packets exhibit a spatial asymmetry, due to the fact that the centers of the applied pulse  and of the Rashba barrier do not coincide. 

The shape of the spin-polarised wavepacket can thus be modified by applying the electromagnetic pulse in suitable locations with respect to the Rashba barrier.

\section{Discussion}
\label{sec-6}
In the previous sections we have discussed the effects of Rashba interaction on helical edge states and its interplay with the photoexcitation due to an en electromagnetic field. This section is devoted to some aspects that have not been addressed so far. In the first instance, we have focussed on the effects of the electric field $E(x,t)$, while the Zeeman coupling due to the magnetic field has not been considered. Secondly,    our analysis has been carried out by assuming that electrons are independent, neglecting  electron-electron interaction, whose effects deserve a comment. Finally, we would like to sketch possible setup realisations where the predicted effects may be detected.

\subsection{Effects of Zeeman coupling}
The time-dependent electric field of the applied pulse necessarily involves also a magnetic field, which in turn generates a Zeeman coupling to the edge states. In particular, a Zeeman field that is orthogonal to the native spin quantisation axis of the edge states, is known to cause a magnetic gap $\Delta_Z$ on the otherwise linear helical spectrum~[\onlinecite{fu-kane_2009}].
We shall now argue that such effect is actually negligible in the situations that are relevant to our analysis. 

In the first instance we  observe that the gap can be made vanishing by suitably choosing the 
polarisation of the electromagnetic radiation: in HgTe/CdTe, for instance, the bulk inversion asymmetry terms make the spin quantisation axis of the helical edge states lie on the quantum well plane, perpendicular to the edge direction~[\onlinecite{qi_2008},\onlinecite{zhang_PRB_2010}]. Thus, if $B$ is aligned along such direction and $E$  along the edge boundary no  gap arises at all.  

Secondly, one can estimate the maximal magnetic gap, corresponding to ``worst case'' scenario of a Zeeman field   orthogonal to the spin quantisation direction. From Maxwell's equation one has $\delta B/ \delta x =  \delta E /c^2  \delta t$~[\onlinecite{nota-system}]. If one assumes that the electric and magnetic field amplitudes, localised region of size $\Delta$,  ramp  up over a timescale $\tau$ from $0$ to  maximal values $E_{max}$ and $B_{max}$, respectively,   one can estimate $B_{max} \simeq E_{max} \Delta/ c^2 \tau$. In particular, taking e.g. $E_{max}=1 {\rm kV/m}$ for an electric pulse applied for $\tau \sim 100 \, {\rm fs}$ over a length scale $\Delta\sim 100\,{\rm nm}$, one obtains
$
B_{max} \sim   10 \, {\rm nT}
$.
The related maximal Zeeman energy is $\Delta_Z=g \mu_B B_{max}$, where $\mu_B$ is Bohr magneton and the $g$-factor is estimated to be $g \sim 0.5$ for InAs/GaSb quantum wells~[\onlinecite{knez_2014}] and $g\sim 20$ for HgTe/CdTe quantum wells~[\onlinecite{molenkamp-zhang_jpsj}]. Then, the maximal magnetic gap is $\Delta_Z \sim 10^{-8} \, {\rm meV}$, which is extremely small when compared to the bulk gap. The omission of this effect, assumed at the beginning of the paper, is thus legitimate.

\subsection{Effects of electron-electron interaction}
In one-dimensional systems electron-electron interaction is typically expected to have dramatic effects, often leading to Luttinger liquid physics~[\onlinecite{zhang_PRL_2006}]. However, it should be mentioned that, despite the large number of theoretical studies concerning helical Luttinger liquids (HLL),  no clear experimental evidence of HLL physics has been observed in the edge states of two-dimensional TIs yet. Presently, the only exception is probably the work of Ref.[\onlinecite{du2015}], where the suppression of conductance at low temperature and low bias
voltage seen in the InAs/GaSb edge states was attributed to strong electron-electron interaction, although a different interpretation in terms of weakly interacting electrons coupled to local magnetic moments was recently put forward~[\onlinecite{glazman_2016}]. 
It is thus still questionable whether electron-electron interaction actually plays a significant role in the helical edge states. Yet, it is worth outlining how the scenario described so far would be modified. 

To this purpose, we now include a density-density interaction, 
$
\hat{\mathcal{H}}_{int} = \hat{\mathcal{H}}_{int,4}\, \, +\,\, \hat{\mathcal{H}}_{int,2}
$, 
where 
$
\hat{\mathcal{H}}_{int,4} = \sum_{\sigma=\uparrow,\downarrow}\iint dx_1 dx_2 \,g_4(|x_1-x_2|)     \hat{n}_\sigma(x_1)   \, \hat{n}_\sigma(x_2)/2$ and $
\hat{\mathcal{H}}_{int,2} = \iint dx_1 dx_2 \, g_2(|x_1-x_2|)  \, \hat{n}_\uparrow(x_1)  \, \hat{n}_\downarrow(x_2)$ 
describe the screened intra-spin and the inter-spin  interaction, respectively.
The whole interaction term can be equivalently rewritten as
\begin{eqnarray}
\hat{\mathcal{H}}_{int} &= & \!\frac{1}{4} \iint \!dx_1 dx_2\, (g_4+g_2)(|x_1-x_2|)    \,  \hat{n}(x_1) \,\hat{n}(x_2)   \,\,+\,\,     \nonumber \\
&+ & \!\frac{1}{4} \!\iint \!dx_1 dx_2\, (g_4-g_2)(|x_1-x_2|)  \hat{s}_3(x_1) \hat{s}_3(x_2) ,\,\,\hspace{0.5cm} \label{Hint-def}
\end{eqnarray}
where $\hat{n}=\hat{n}_\uparrow +\hat{n}_\downarrow$ is the local density, and $\hat{s}_3 \doteq \hat{n}_\uparrow -\hat{n}_\downarrow$ is the local third component of  spin, up to a $\hbar/2$ factor. 
The first term on the right-hand side of Eq.(\ref{Hint-def}), proportional to the sum $g_4+g_2$ of the coupling constants, is the SU(2)-symmetric interaction term, for it commutes with all three total spin components $\hat{S}_i= \hbar  \int \Psi \sigma_i \Psi/2$ ($i=1,2,3$). In contrast, the second term in Eq.(\ref{Hint-def}), proportional to the difference $g_4-g_2$ of the coupling constants, reduces the symmetry to a U(1), for it commutes only with the third component $\hat{S}_3=\hbar  \int (\hat{n}_\uparrow -\hat{n}_\downarrow)/2$. 

The effects of the interaction are typically accounted for by adopting the bosonization formalism~[\onlinecite{vondelft}], where fermionic operators are expressed as vertex operators of two bosonic fields $\Phi$  and $\Theta$  through $\psi_{\uparrow,\downarrow}=\exp[i\sqrt{\pi}(\Theta \pm \Phi)]/\sqrt{2\pi a}$, with $a$ denoting the short-distance cutoff.  
In bosonization language the density $\hat{n}$ appearing in the first term of Eq.(\ref{Hint-def}) is linear in the derivative $\partial_x\Phi$ of the bosonic field~$\Phi$. In contrast, the $\hat{s}_3=\hat{n}_\uparrow -\hat{n}_\downarrow$ component appearing in the second term of Eq.(\ref{Hint-def})  must be considered with care. When  Rashba interaction is absent, $\hat{s}_3$ coincides with the current (up to a prefactor $v_F$), due to the helical nature of the electronic states, and it is expressed  in bosonization as the derivative $\partial_x \Theta$ of the dual field~$\Theta$. As a consequence, both   interaction terms in Eq.(\ref{Hint-def}) can be recast into Luttinger liquid forms, quadratic in $\partial_x \Phi$ and $\partial_x \Theta$, respectively. 

However, in the presence of a Rashba interaction, $\hat{s}_3=\hat{n}_\uparrow -\hat{n}_\downarrow$ does {\it not} describe the current, which is given by Eq.(\ref{J-def}) instead.  
For these reasons, $\hat{s}_3$ is   not linear in the bosonic field~$\partial_x \Theta$ and its   presence, related to the difference $g_4-g_2$ in the second  term of Eq.(\ref{Hint-def}), cannot be expected to be harmless when Rashba interaction is present. In order to evaluate its impact beyond a perturbative approach, it is worth rewriting the interacting part $\hat{\mathcal{H}}_{int}$ in terms of the chiral fields $\chi_\pm$, in which the  single particle  Hamiltonian $\hat{\mathcal{H}}_\circ+\hat{\mathcal{H}}_{em}$ (linear band+Rashba+electromagnetic coupling) is diagonal. Using the transformation (\ref{Psiprime-Psi}) to the chiral fields, one has 
\begin{equation}
\hat{n}=\hat{n}_\uparrow +\hat{n}_\downarrow  \, =\, \Psi^\dagger  \sigma_0  \Psi = {\rm X}^\dagger \sigma_0  {\rm X}^{}  \label{dens-transf} 
\end{equation}
and
\begin{eqnarray}
\hat{s}_3= \hat{n}_\uparrow -\hat{n}_\downarrow  \, &=&\, \Psi^\dagger  \sigma_3 \Psi \, = \,   {\rm X}^\dagger (\cos\theta_R \sigma_3-\sin\theta_R  \sigma_2){\rm X}^{} =  \nonumber \\
  &=& (\hat{J} \cos\theta_R \, - \, \hat{J}^{\rm BS}  \sin\theta_R)/v(x) \,  .\label{curr-transf}
\end{eqnarray}
Here $\hat{J}$ is the current, defined in Eq.(\ref{J-def}), which was shown to be purely diagonal in the chiral basis [see Eq.(\ref{J-chi-fields})] and thus describes a ``forward scattering'' process, whereas 
\begin{eqnarray}
\hat{J}^{\rm BS}(x)  &\doteq  & 
 v(x) \, {\rm X}^\dagger \sigma_2    {\rm X} =  -i  v(x)  \left( \chi^\dagger_{+} \chi^{}_{-}   -  \chi^\dagger_{-} \chi^{}_{+} \right)  \label{JBS-def} \hspace{0.6cm}
\end{eqnarray}
describes the ``backward scattering'' current, purely off-diagonal  in the chiral basis: its expectation value determines the amplitude for   right-moving electrons to be scattered into left-movers and viceversa.  
Equations (\ref{dens-transf}) and (\ref{curr-transf}) show that, while the density is covariant when passing from the spin basis to the chiral basis, the current is not.

We are interested in describing the physics at lengthscales longer than the typical range of the screened interaction, so that the latter will be henceforth approximated with a local potential $g_{2,4}(|x_1-x_2|) \simeq g_{2,4}\delta(x_1-x_2)$, although this assumption is not strictly necessary. Then, the full Hamiltonian   $\hat{\mathcal{H}}=\hat{\mathcal{H}}_\circ+\hat{\mathcal{H}}_{em}+\hat{\mathcal{H}}_{int}$  is rewritten as
\begin{eqnarray}
\hat{\mathcal{H}} &=& \int dx \,{\rm X}^\dagger\left[ \frac{1}{2}\left\{ v(x)    ,\, {p}_x  -\frac{{\rm e}}{c}  A  \right\} \sigma_3 \, +{\rm e}   V  \sigma_0 \right] {\rm X}^{} + \nonumber\\
& &+   \int dx \, 
 \left\{ \, \,\frac{g_4+g_2}{4}    \, \,   \hat{n}^2(x)   +  \right.   \label{H-new}\\
& & +\left. \frac{g_4-g_2}{4 v^2(x)}  \left[  \hat{J}(x) \cos \theta_R(x) \, - \hat{J}^{\rm BS}(x) \sin\theta_R(x) \,\right]^2  \right\} \,.\nonumber 
\end{eqnarray} 
 
\noindent The first line of Eq.(\ref{H-new}) is the single-particle Hamiltonian   $\hat{\mathcal{H}}_\circ+\hat{\mathcal{H}}_{em}$ describing massless Dirac fermions with the inhomogeneous velocity (\ref{v(x)-def}) encoding the Rashba interaction, coupled to the electromagnetic field. The second line describes a total density-density interaction, whereas the third line involves forward and backward current terms.  
The expression (\ref{H-new}) of the Hamiltonian  in the chiral basis, which is obtained   non-perturbatively in the Rashba coupling~$\alpha_R$, can now be rewritten by bosonizing the chiral fields as 
$
\chi_\pm(x)= \exp[i \sqrt{\pi}(\Theta_\chi \pm \Phi_\chi)(x)]/\sqrt{2\pi a(x)}
$, where $a(x)\doteq \hbar v(x)/E_c$ is a space-dependent short-distance cut-off, expressed in terms of one ultraviolet energy cut-off  $E_c$ and the local velocity $v(x)$, while $\Phi_\chi$ and $\Theta_\chi$ are bosonic fields fulfilling $[\Phi_\chi(x), \partial_y \Theta_\chi(y)]=-i\delta(x-y)$~[\onlinecite{nota-infrared-cutoff}]. Two different scenarios can emerge.

When $g_2=g_4$, the Hamiltonian~$\hat{\mathcal{H}}$ consists of the first two lines of  Eq.(\ref{H-new}) only and, when rewritten in terms of the bosonic field $\Phi_\chi$ and $\Theta_\chi$,  reads $\hat{\mathcal{H}}=\hat{\mathcal{H}}_{ILL}+\hat{\mathcal{H}}_{em}$, where  $\hat{\mathcal{H}}_{ILL}=(\hbar/2)\int dx \,v(x) [(\partial_x\Theta_\chi)^2+(\partial_x\Phi_\chi/K(x))^2]$ is an inhomogeneous Luttinger liquid (ILL)~[\onlinecite{safi1995,maslov1995,ponomarenko1995,dolcini2005,pugnetti_2009,sassetti-dolcetto2015}], characterized by a space-dependent interaction parameter,
\begin{equation}\label{K-Lut}
K(x)=\left(1+\frac{g_4+g_2}{2\pi \hbar v(x)}\right)^{-1/2}\quad,
\end{equation}
that includes the Rashba coupling via the velocity (\ref{v(x)-def}). 
 Electron-electron interaction induces a  non-analytical behavior in the correlation functions (\ref{corr-0}) of the chiral fields~$\chi_\pm$, which combine  in a non-linear way to identify new quasi-particles carrying a non-integer charge~[\onlinecite{kane-fisher_1994}]. Furthermore the electromagnetic field leads to additional phases,   similarly to Eqs.(\ref{chipm-sol}), where the retardation effects encoded in Eq.(\ref{phipm}) are  affected by the Luttinger parameter (\ref{K-Lut}), though~[\onlinecite{levitov_2006},\onlinecite{pugnetti_2009},\onlinecite{dolcini2012}]. As a consequence, the propagation velocity of the photoexcited densities is  typically increased by interaction. 
Despite these modifications, no single particle backscattering arises, i.e.  the property $\langle \chi^\dagger_{+} \chi^{}_{-}\rangle = 0$ still holds for a ILL.  This result generalises the topological protection, found in Ref.[\onlinecite{foster_2016}] for $g_2=g_4$, to the additional presence of the electromagnetic field.

However, when $g_2 \neq g_4$, the contribution from the third line of Eq.(\ref{H-new}) leads to new features, as shown by an inspection of the square therein. On the one hand  ``forward-forward''~$\hat{J}$-$\hat{J}$  terms appear that, once bosonized, are quadratic in the bosonic field $\partial_x\Theta_\chi$. These terms can again be included into the ILL Hamiltonian, and  modify the profiles of both the velocity $v(x)$ and the Luttinger parameter $K(x)$ through its Rashba coefficient $\cos^2\theta_R(x)$. On the other hand, however, the interplay of the Rashba interaction ($\sin\theta_R \neq 0$)  with the $g_4-g_2$ interaction difference also leads to  ``forward$\times$backward'' $\hat{J}$-$\hat{J}^{\rm BS}$ terms and  to ``backward$\times$backward'' $\hat{J}^{\rm BS}$-$\hat{J}^{\rm BS}$   terms, expressed as  $\partial_x \Theta_\chi \cos[\sqrt{4\pi} \Phi_\chi]$ and $\cos[\sqrt{16\pi} \Phi_\chi]$ in the bosonization language, respectively. 
The former type of terms  has been investigated   in the case of weak Rashba interaction and vanishing electromagnetic field, and a  renormalisation group analysis has shown that it leads to two-particle backscattering processes that modify the temperature dependence of the conductance~[\onlinecite{johannesson_2010,crepin_2012,schmidt_2012,geissler_2014,mirlin_2014,geissler_2015}]. The latter type of terms, which is of order $\mathcal{O}(\alpha_R^2)$, describe   the umklapp scattering. Such processes become  important for commensurate filling $k_F=\pi/2a$, and their effects depend  on the   spatial extension of the Rashba interaction: in particular, for an extended Rashba coupling  a gap can open up, whereas for a localised Rashba impurity transmission is possible,  with backscattering though~[\onlinecite{zhang_PRL_2006}].

\subsection{Possible setup realisations}
Let us now discuss possible   setups where the effects of the interplay between Rashba interaction and electromagnetic coupling may be  observed. Two main  realisations of two-dimensional TIs presently exist, namely   in HgTe/CdTe
[\onlinecite{bernevig_science_2006,konig_2006,molenkamp-zhang_jpsj,roth_2009,brune_2012}] and in InAs/GaSb~[\onlinecite{liu-zhang_2008,knez_2007,knez_2014,spanton_2014}] quantum wells. In their topological phase, conducting helical edge states  appear and exhibit a linear dispersion with a Fermi velocity $v_F \simeq 5 \times 10^5 {\rm m/s}$ and $v_F \simeq 2 \times 10^4 {\rm m/s}$,  respectively~[\onlinecite{molenkamp-zhang_jpsj},\onlinecite{knez_2014}],   within a bulk gap   $E_g \sim 30 \,{\rm meV}$. The phase breaking length $L_\phi$, i.e. the length scale for which the  analysis carried out in this paper is valid, is of the order of a few micrometers at Kelvin temperatures. 

A Rashba barrier can be created by deforming locally the geometry of a quantum well boundary, thereby inducing a local strain on the edge states. This can be done, for instance, by lateral etching  a re-entrance on the boundary, similarly to what has been proposed in various works about inter-edge tunneling setups~[\onlinecite{teo-kane2009,chamon2009,trauz-recher,richter_2011,dolcini2011,dolcetto-sassetti2012,citro-sassetti,sassetti-ferraro2013,sternativo_2014,dolcetto2014,dolcini2015}], or by growing a vertical ``bump'' of circular arcs on the quantum well~[\onlinecite{ojanen_2011}]. In both cases, the curvature of the deformation and the dc voltage $V_g$  applied to a wedge gate enable  one to tune the strength of the Rashba coupling. 

\begin{figure}[h]
\centering
\includegraphics[width=\columnwidth]{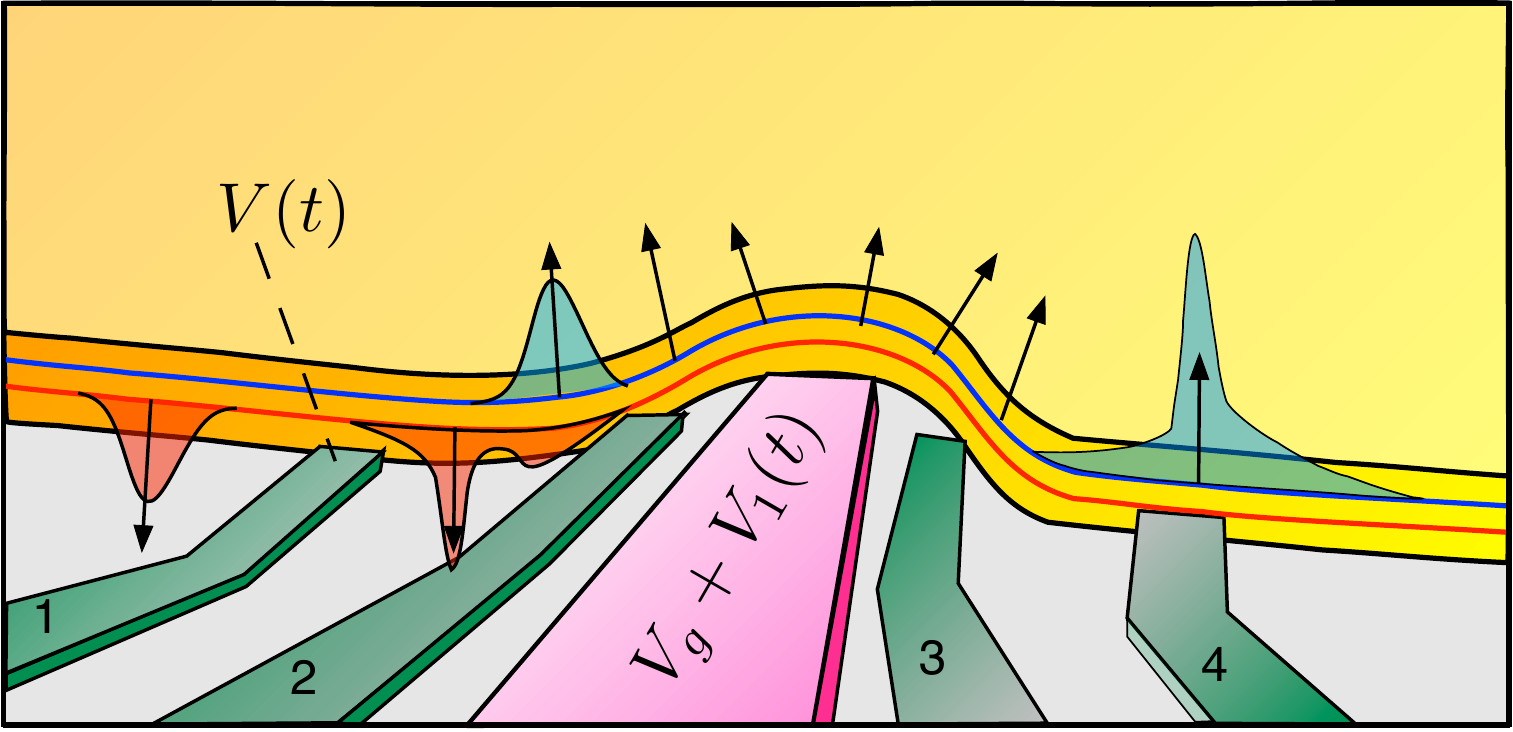}
\caption{
Possible realisation scheme of a setup for the photoexcitation in the presence of Rashba interaction. A re-entrance etched on one edge of the quantum well in the topological phase and a dc voltage $V_g$ applied to a wedge gate enable one to control the Rashba interaction. Electromagnetic pulses $V(t)$ applied to finger gates and/or the the wedge gate photoexcite electrons along the edge states. The different scenarios illustrated in Figs.~\ref{Fig-dens-1}, \ref{Fig-dens-2} and \ref{Fig-dens-3} can be realised, depending on the  choice of the ac biased gate, the strength  of the Rashba interaction, the size of the finger gates and the duration of the applied pulse. The shape of the spin-polarized photoexcited wavepackets and their propagation timescales can thus be controlled, as discussed in the text.}
\label{Fig-setup}
\end{figure}
Furthermore, as observed in the Introduction, a localised electromagnetic pulse can be generated with two techniques. 
The first one is the use of near field scanning optical microscopy~[\onlinecite{novotny_review,koch1997,novotny2003,nomura2011,nomura2015}] operating in the illumination mode: an optical fiber with a thin aperture of tens of nanometers, positioned near the edge, excites a strong electric field at the tip apex~[\onlinecite{koch1997,novotny2003,nomura2011,nomura2015}]. With this sophisticated technique one obtains  localised pulses, whose spatial center can also be easily be displaced, so that all situations described in Sec.\ref{sec-5} can be achieved, from photo excitation occurring away from the Rashba barrier to the case of its overlap  with the Rashba barrier. 
 The second approach to generate a localised electromagnetic pulse is somewhat simpler and more straightforward: It amounts to utilising side finger gate electrodes, deposited close to a boundary of the QSH bar  and biased by   time-dependent voltages experienced by the electrons in the edge, similarly to what has been proposed for a 2DEG~[\onlinecite{levitov_2006,glattli_PRB_2013,glattli2013,bocquillon2014}]. In this case the spatial extension of the electric pulse is determined by the lateral width of the finger electrode,  $\sim 100 {\rm nm}$. Note that  the pure photoexcitation process does not involve any electron tunneling from the finger electrodes, differently from the case of electron pumps~[\onlinecite{feve2007,ritchie_2007,feve2011,kataoka2013,kataoka2015,janssen2016}]. A possible setup scheme is sketched in Fig.~\ref{Fig-setup}.  In particular, the situation described in Sec.\ref{sec-5-a} and illustrated in Fig.~\ref{Fig-dens-1}, where the photoexcitation and the Rashba region are spatially separated, can be realised by applying an electromagnetic pulse  $V(t)$ to a finger electrode away from the geometrical re-entrance.  
In contrast, the situations of interplay between photoexcitation and Rashba coupling, described in Secs.\ref{sec-5-b} and \ref{sec-5-c} and illustrated in Figs.~\ref{Fig-dens-2} and \ref{Fig-dens-3}, can be implemented by applying an additional ac bias to the electrode into the wedge of the etched re-entrance mentioned above.  With the proposed setup the shape of the photoexcited wave packets can be tailored by the parameters determining the Rashba barrier, i.e. the curvature of the geometrical re-entrance and the value~$V_g$ of the wedge gate voltage, by the size $\Delta$ of the finger gates, and  by the amplitude $E_0$ and the   duration $\tau$ of the applied pulse. 

A comment about $\tau$ is in order. The rule of thumb dictated by time-dependent perturbation theory would require that $\tau>\hbar/E_g$ in order to avoid transition across the bulk gap $E_g$. However, the actual transition rate also depends on the field amplitude. In this respect, the advantage of the setup depicted in Fig.~\ref{Fig-setup} is that the finger gate electrodes are localised at the boundary of the quantum well, so that the electric field is applied on the edge states {\it directly}, while the bulk states experience a strongly reduced amplitude. This suppresses the role of transitions between bulk states. Still, consistency requires that the energy imparted by the electric field onto the edge state  electrons must be smaller than the bulk energy gap. An estimate can be obtained e.g. by inspecting the energy range over which the momentum distributions $f_\pm(k,t)$ change with respect to the equilibrium case, which roughly leads to the condition $\tau < E_g/({\rm e} E_0 v_F)$, where $E_0$ is the amplitude of the field experienced by the edge states~[\onlinecite{dolcini_2016}].

The recent progress in pump-probe experiments and photo-current spectroscopy~[\onlinecite{lesueur2009,glattli2013,bocquillon2014,holleitner2014,holleitner2015,jarillo-herrero2016}], make the time-resolved  detection of the photoexcited wave packets realistically accessible nowadays.
These results seem promising in view of utilising  two-dimensional TIs as possible alternative platform for an electron quantum optics~[\onlinecite{bercioux_2013,buttiker_2013,recher_2015,sassetti_2016}], which is nowadays mostly implemented in quantum Hall systems~[\onlinecite{glattli2013},\onlinecite{bocquillon2014}] with the unavoidable limitation of the need for strong magnetic fields. Time-reversal TIs, which are based on spin-orbit coupling, are immune to such drawback and offer the additional possibility of generating spin-polarised electron wave packets.


\section{Conclusions}
\label{sec-7}
In this paper we have analyzed non-perturbatively the interplay between the Rashba interaction and the electromagnetic field in the helical states flowing at an edge of a two-dimensional topological insulator. By applying a local rotation to a chiral basis ${\rm X}$, Eq.(\ref{Psi-from-X-compact}), we have shown that the problem is equivalent to a system of massless Dirac fermions, Eq.(\ref{Hchi}), propagating with an inhomogeneous velocity that turns out to be always enhanced by the Rashba interaction profile, as compared to its bare Fermi velocity value $v_F$ [see Eq.(\ref{v(x)-def})]. Such mapping unveils important physical aspects that we have discussed in detail.  

In Sec.\ref{sec-3} we have first addressed the case without electromagnetic field. While the customary spin basis~$\Psi$ identifies right- and left-moving electrons only far from a Rashba interaction region,   the two components $\chi_{+}$ and $\chi_{-}$ of the chiral basis~${\rm X}$ have been shown to describe genuine right- and left-movers even in the presence of Rashba interaction, and can thus be considered the natural basis for the problem. Introducing a general model for the Rashba profile [see Eq.(\ref{alphaR-region}) and Fig.~\ref{Fig-Rashba-profile}], we have shown the behavior of the electron wavefunction propagating e.g. rightwards towards a Rashba region with a purely spin-$\uparrow$ component: By approaching the Rashba region the wavefunction increases its spatial period and displays an additional spin-$\downarrow$ component, emerging over a lengthscale related to the smoothening length of the Rashba profile (see Fig.~\ref{Fig-wave}). 
 
In Sec.\ref{sec-4} we have then included the electromagnetic field. We have shown that, although the fields $\chi_\pm$ modify their character of right- and left-movers, they remain dynamically decoupled, implying that no backscattering occurs, despite the inelestic and time-reversal breaking processes induced by the electromagnetic field. Furthermore, we have explicitly derived   the photoexcited electron densities $\Delta n_\pm$ [see Eqs.(\ref{Delta-n_+}) and (\ref{Delta-n_-})], which turn out to be expressed as   a convolution of the applied electric field where   Rashba interaction appears as a retardation effect. Furthermore, this result enabled us to generalize the chiral anomaly effect, usually discussed in the case of massless Dirac fermions with constant velocity, to the case of fermions with inhomogeneous velocity realized by the presence of the Rashba interaction [see Eqs.(\ref{chi-ano-2-a}) and (\ref{chi-ano-2-b})].
 
In Sec.\ref{sec-5} we have then applied these general results to the example of electron photoexcitation due to a Gaussian electric pulse, localized over a lengthscale $\Delta$ and applied for a timescale $\tau$, in the presence of a Rashba barrier extending over a lengthscale $W_R$ along the edge. We have shown that, when the photoexcitation occurs far from the Rashba barrier (see Fig.~\ref{Fig-dens-1}) the latter acts as a ``superluminal gate'', i.e. it boosts the impinging wavepacket  into space-time regions that are beyond the light-cone dictated by the bare Fermi velocity $v_F$. In contrast, when the same Gaussian pulse is overlapping the Rashba interaction region, the asymptotically emerging wavepackets turn out to be squeezed. In particular the profile depends on the relative extension and duration of the applied pulse as compared to the Rashba region (see Figs.~\ref{Fig-dens-2} and~\ref{Fig-dens-3}). 

Finally, in Sec.\ref{sec-6} we have shown that the Zeeman coupling plays a minor role in the problem, and we have discussed the role of electron-electron interaction, considering both cases of intra-spin ($g_4$) and inter-spin ($g_2$)  density-density couplings [see Eq.(\ref{Hint-def})]. We have shown that  for $g_2=g_4$ the system can be mapped into an inhomogeneous Luttinger liquid with interaction parameter (\ref{K-Lut}) and the topological protection (no backscattering) is robust to both interactions and the electromagnetic field. In contrast, for $g_4 \neq g_2$ an interplay between Rashba and electron-electron interaction occurs [see Eq.(\ref{H-new})] that can lead to two-particle backscattering and/or to umklapp terms. Finally, we have have discussed a possible realisation of a setup (see Fig.~\ref{Fig-setup}), where the effects analysed here may possibly be observed. 

These results suggest that the Rashba interaction, which are often regarded to as an unwanted disorder effect, may be utilized in the near future to tailor the shape of spin-polarized photoexcited wavepackets or to control their propagation timescales, paving the way to a QSH based electron quantum optics.

\acknowledgments
Illuminating discussions with J. C. Budich, F. Cr\'epin, F. Geissler, A. Montorsi, P. Recher, F. Rossi and B.~Trauzettel are greatly acknowledged. \\

\appendix
\section{Ballistic flight-time for the Rashba region model}
\label{AppA}
The   ballistic flight-time, defined in Eq.(\ref{s-def}), can be given an analytical expression in the case of the model (\ref{alphaR-region}) proposed for the Rashba interaction profile. In such case  it is natural to choose the center $x_R$ of the Rashba profile as the fixed reference point $x_r$ appearing in the definition (\ref{s-def}). A lengthy but straightforward calculation  leads to obtain 
\begin{eqnarray}
s(x) &= &\frac{1}{v_F} \left\{ x-x_R+\mbox{sgn}(x-x_R)\frac{l_R}{2} \, \,\times \right.  \label{s-model}  \\
& & \times \left. \left[ \ln \frac{A(x-x_R)}{A(0)}-\frac{1}{\sqrt{1+G_R^2}}  \ln \frac{B(x-x_R)}{B(0)}\right] \right\}\nonumber
\end{eqnarray}
where
\begin{eqnarray}
A(x)& \doteq &e^{-\frac{|x|}{l_R/2}}+e^{-\frac{W_R}{l_R}}+\\
& & +\sqrt{\left(e^{-\frac{|x|}{l_R/2}}+e^{-\frac{W_R}{l_R}}\right)^2+G_R^2 \, e^{-4\frac{|x|}{l_R}}} \quad,\nonumber 
\end{eqnarray}
\begin{eqnarray}
B(x)& \doteq &e^{-\frac{|x|}{l_R/2}}(1+G_R^2)+e^{-\frac{W_R}{l_R}}+\sqrt{ 1+G_R^2} \times \nonumber \\
& & \times \sqrt{ \left(e^{-\frac{|x|}{l_R/2}}+e^{-\frac{W_R}{l_R}}\right)^2+G_R^2 \, e^{-4\frac{|x|}{l_R}} }
\end{eqnarray}
and
$
G_R \doteq g_R  \,[1+\exp(-W_R/l_R)]
$. 
In particular, in the limit $\lambda_R \ll W_R$ of the Rashba ``barrier'', Eq.(\ref{s-model}) simplifies to the piecewise linear expression
\begin{eqnarray}
\lefteqn{ s(x) \simeq \frac{\mbox{sgn}(x-x_R)}{v_F} \,\times } & & \label{s-model-box}\\
& & \times \left\{ 
\begin{array}{lcl}   |x-x_R| -\frac{W_R}{2} \left(1- \frac{1}{\sqrt{1+g_R^2}}\right)   & \mbox{for } & |x-x_R|> \frac{W_R}{2} \\
\frac{|x-x_R|}{\sqrt{1+g_R^2}}  & \mbox{for } & |x-x_R| <   \frac{W_R}{2}
\end{array} \right. . \nonumber
\end{eqnarray}


\begin{thebibliography}{99}
\bibitem{nota-dimensione-phi} Note that, since the energy quantum number $E$ is  a continuous variable, the  dimension of the wave functions $\varphi^{\pm}_{\uparrow,\downarrow}$  is $\mbox{(space $\times$ energy)}^{-1/2}$.
\bibitem{nota-barrier} We adopt the term ``barrier'' because in the limit $l_R \ll W_R$ the profile is reminiscent of the typical barrier potential in quantum mechanics. The hyphenation is because the Rashba interaction alone cannot actually backscatter electrons, as observed in the Introduction.
\bibitem{nota-system} For this estimate it is useful to temporarily switch to the MKS unit system, where the more customary ${\rm meV}$ can be used for the electric potential.
\bibitem{nota-infrared-cutoff} There is also an infrared energy cut-off $E_L$, related to the length $L$ of the system, which for an inhomogenous system  reads $E_L=2\pi \hbar/\int_{-L/2}^{L/2}  dx^\prime v^{-1}(x^\prime)$.
\end{thebibliography}
\end{document}